\documentclass[11pt,a4paper,final]{article}
\usepackage{jheppub}
\usepackage[latin1]{inputenc}

\usepackage{bbm,bm}
\usepackage{graphicx}
\usepackage{wrapfig}
\usepackage{amssymb}
\usepackage{mathrsfs}
\usepackage{amsmath}
\usepackage{comment}
\usepackage{amsbsy}
\usepackage{mathrsfs}
\usepackage{amsfonts}
\usepackage{epsf,color,colordvi,pifont}
\usepackage{lineno}%(enumera las lineas)
\usepackage{nicefrac}
\usepackage[notref,notcite]{showkeys} %% just for draft versions
\def\deltabar{{\mathchar'26\mkern-9mu\delta}}

\DeclareFontFamily{OT1}{pzc}{}
\DeclareFontShape{OT1}{pzc}{m}{it}{<-> s * [1.10] pzcmi7t}{}
\DeclareMathAlphabet{\mathpzc}{OT1}{pzc}{m}{it}
\DeclareMathOperator\arctanh{arctanh}

\usepackage{url}

\title{Light dark matter candidates in intense laser pulses II: the relevance  of the spin degrees of freedom}

\author{S.  Villalba-Ch\'avez}

\author{and C.  M\"{u}ller}

\affiliation{Institut f\"{u}r Theoretische Physik I, Heinrich-Heine-Universit\"{a}t D\"{u}sseldorf\\ Universit\"{a}tsstr. 1, 40225 D\"{u}sseldorf, Germany}

\emailAdd{selym@tp1.uni-duesseldorf.de}
\emailAdd{c.mueller@tp1.uni-duesseldorf.de}

\abstract{Optical searches assisted  by the  field  of a   laser pulse   might allow for  exploring  a variety of  not yet  detected  dark matter 
candidates such as  hidden-photons  and  scalar minicharged particles.  These hypothetical degrees of freedom  may be understood as a natural consequence 
of extensions  of the Standard  Model   incorporating  a  hidden  $\rm U(1)$-gauge  sector. In this paper, we study the effects induced  by both  candidates on the 
propagation of a probe electromagnetic waves in the vacuum polarized by a long laser pulse of moderate intensity, this way complementing our previous study [JHEP \textbf{06}, $177$ ($2015$)]. 
We describe how  the absence of a spin in the scalar charged carriers  modifies the photon-paraphoton oscillations as compared with a fermionic minicharge  model. 
In particular, we find that the regime close to their lowest threshold mass  might provide the most stringent upper limit for minicharged scalars. The pure-laser based  experiment   investigated here  could allow 
for excluding  a sector in the parameter space of the particles which  has not been experimentally ruled out by setups  driven by dipole magnets. We explain how the 
sign of the ellipticity and rotation of the polarization plane  acquired by a probe photon--in combination with their dependencies on the pulse parameters--can be exploited 
to elucidate the quantum statistics of the charge carriers.
}

\keywords{Beyond the Standard Model, Minicharged Particles, Hidden Photons,  Laser Fields.}

\begin{document}
\maketitle
\flushbottom

%%%%%%%%%%%%%%%%%%%%%%%%%%%%%%%%%%%%%%%%%%%%%%%%%%%%%%%%%%%%%%%%%%%%%%%%%%%%%%%%%%%%%%%%%%%%%%%%%%%%%%%%%%%%%%%%%%%%%%%%%%%%%%%%%%%%%%
\section{Introduction}
%%%%%%%%%%%%%%%%%%%%%%%%%%%%%%%%%%%%%%%%%%%%%%%%%%%%%%%%%%%%%%%%%%%%%%%%%%%%%%%%%%%%%%%%%%%%%%%%%%%%%%%%%%%%%%%%%%%%%%%%%%%%%%%%%%%%

Identifying the dark matter in the Universe and consistently incorporating it into the Standard Model (SM) constitute challenging problems in today's particle physics. Cosmological  as well as astrophysical observations  
provide substantial evidence that only a small fraction $4,5\%$ of matter is made out of the elementary building blocks of the SM but there is not yet a clear  idea about the origin and nature of the dark  matter 
\cite{Hinshaw:2012aka,Clowe:2006eq,Iocco:2015xga,Ade:2015xua}.  This fact  evidences why the SM is currently accepted as an effective theory which must be embedded into a more general framework at higher energy scales. Such 
an  enlarged theory is expected to offer us a comprehensive theoretical  understanding  about a variety of  central problems including the charge quantization, which presently lacks  an experimentally verifiable explanation.  
While some extensions of the SM provide mechanisms for enforcing charge quantization,  other  scenarios including carriers of small unquantized charge are not discarded. Indeed, effective theories  containing an extra $\rm U(1)$ 
gauge field \cite{Witten:1984dg,Lebedev:2008un,Lebedev:2009ag,Goodsell:2009xc} kinematically mixed with the electromagnetic sector \cite{Okun:1982xi,Langacker:2008yv,Ahlers:2007rd,Ahlers:2007qf}, introduce this sort of 
Mini-Charged Particles (MCPs) \cite{Holdom:1985ag,Dobrescu:2006au,Gies:2006ca} in a natural way. The fact that at low energies these carriers are not observed might be considered as an evidence indicating  that the sector to 
which they belong  interacts only very weakly with the well established SM branch.  It is, in addition, reasonable  to assume  that a hypothetical  existence of MCPs induces nonlinear interactions in the electromagnetic field 
provided they are  very light sub-eV particles  minimally coupled to the ``visible'' $\rm U(1)$ sector \cite{Dobrich:2009kd,Gies:2008wv}. Slight discrepancies are expected then as compared with the inherent phenomenology  of Quantum Electrodynamics (QED).  Indeed, motivated  by this possibility, various  experimental 
collaborations   have  imposed constrains  and ruled out sectors in the parameter space of these hypothetical  degrees  freedom. 

The  phenomena of interest which have been  exploited in this research area  so far are summarized in several reviews~\cite{Jaeckel:2010ni,Ringwald:2012hr,Hewett:2012ns,Essig:2013lka}. These searches fall into two categories 
depending upon the scenario under consideration.  On the one side, there are searches  relying on astro-cosmological observations. They  provide  the most stringent constraints at present.  Indeed, arguments related to  energy 
loss which is not observed in Horizontal Branch stars, limit the relative charge in MCPs to $\epsilon\lesssim 10^{-14}$ for masses below a few $\rm keV$ \cite{davidson}. However, further investigations in this direction have 
provided arguments indicating the extent to which this bound is sensitive to the inclusion of macroscopic and microscopic parameters of the star, as well as to certain processes that might attenuate it significantly and, simultaneously, 
elude it from  our perception \cite{evading,Masso:2006gc,Masso:2005ym,Jaeckel:2006id}. The described vulnerability  in  the astro-cosmological constraints is a strong motivation for considering, on the other side, well-controlled 
laboratory-based  searches as a complementary approach. Generally, these  have been  conducted through high precision experiments  looking for the  birefringence and dichroism of the vacuum\footnote{Recent theoretical studies on 
the birefringence and dichroism of the vacuum in an external magnetic field  can be found in \cite{Dittrich,Hattori,VillalbaChavez:2012ea}.}  \cite{Cameron:1993mr,Zavattini:2007ee,DellaValle:2013xs,BMVreport,Chen:2006cd},  
modifications in  Coulomb's law \cite{Jaeckel:2009dh,Jaeckel:2010xx} or through the regeneration of photons from a hidden photon field  in  ``Light Shining Through a Wall'' setups \cite{Ehret:2010mh,Chou:2007zzc,Afanasev:2008jt,Pugnat:2007nu,Robilliard:2007bq,Fouche:2008jk}. 
For details, variants and prospects of this kind of experiment we refer also  to  Refs.~\cite{Redondo:2010dp,Arias:2010bh,Gies:2006hv,Dobrich:2012sw,Sikivie:2007qm,Bahre:2013ywa}. Most of these 
experiments require the presence of a static external magnetic field to induce vacuum polarization mediated by virtual pairs of MCPs. As a general rule, the relevant observables  depend on the field strength as well as its spatial 
extend and, usually, such dependencies allow  for finding more  stringent  bounds as both parameters increase. However, today our technical capability in laboratories  are quite limited,  allowing  us to achieve constant magnetic 
fields no higher than $\sim10^5\ \mathrm{G}$ along effective  distances of the order of $\sim 1\ \mathrm{km}$. 

Focused laser pulses of few micrometer extension can produce much stronger magnetic fields but they are inhomogeneously  distributed \cite{Di_Piazza_2012}. For instance, the  highest peak intensity achieved so far  $2\times 10^{22}\ \rm W/cm^2$ 
\cite{yanovsky} corresponds to a magnetic field strength  of $9\times 10^9\ \rm G$. Besides, peak magnetic fields exceeding $\sim 10^{11}\ \rm G$ are likely to be reached by the  ongoing  ELI and  XCELS projects \cite{ELI,xcels},  in  which  
intensities greater than $10^{25}\ \rm W/cm^2$ are  envisaged. In view of these perspectives, high-intensity laser pulses are potential  tools with which  nonlinear phenomena in strong  field QED \cite{DiPiazza:2007yx,nonlin1,Hatsagortsyan:2011bp,King:2014vha} 
can be observed for the first time. Obviously, this would also provide  an  opportunity for detecting the birefrigence of the vacuum \cite{Heinzl}.\footnote{See  \cite{Villalba-Chavez:2013gma,Dinu:2013gaa,Dinu:2014tsa,Villalba-Chavez:2014nya}  for new insights on the vacuum refractive indices in plane-wave fields} 
Indeed, motivated by this idea,  the HIBEF consortium has proposed a laser-based  experiment  which combines a Petawatt optical  laser with a x-ray free electron laser \cite{HIBEF}. Similarly to  setups  driven by static magnetic 
fields,  polarimetric experiments assisted  by an external laser-wave might also constitute a sensitive probe for searching  weakly interacting particles. Although studies of this nature  have been put forward for the case of  
axion-like particles \cite{mendonza,Gies:2008wv,Dobrich:2010hi,Tommasini:2009nh,Villalba-Chavez:2013bda,Villalba-Chavez:2013goa}, the estimate of the exclusion limits for MCPs and hidden photon fields from laser-based 
polarimetric searches is much less developed.  

A first study on MCPs has been  given by the authors in Refs.~\cite{Villalba-Chavez:2013gma}. Later, in part I of this series \cite{Villalba-Chavez:2014nya},  a further step was performed  by investigating the optical effects 
resulting from an extended model containing fermionic  MCPs and a  hidden photon field.  There we revealed that, at moderate intensities $\lesssim 10^{16}$\ \rm W/cm$^2$  as provided by the nanosecond frontends of the  PHELIX 
laser \cite{PHELIX} and LULI system \cite{LULI},  high-precision polarimetric  measurements could improve the existing laboratory  upper bounds for the coupling constant of  MCPs by an order of magnitude for masses of the order of 
$m_\epsilon\sim \rm eV$. However, charge carriers  with unquantized electric charges might  be realized in nature not only as fermions but also as scalar particles \cite{Ahlers:2006iz}. Hence, a complete study of this  subject 
requires  in addition the  insights coming from the polarization tensor  \cite{baier,Mitter,Meuren:2013oya}    that  results from  the Green's function of scalar MCPs and in which the field of the wave  is incorporated in full 
glory. For a monochromatic plane-wave background, corresponding expressions in a pure QED context have been obtained previously \cite{baier,VillalbaChavez:2012bb}. In this paper, we study the effects resulting from  scalar 
minicharges and paraphotons  in a plausible polarimetric setup assisted by a long laser pulse of moderate intensity. We show  how  the absence of spin in the scalar charge carriers  modifies the photon-paraphoton oscillations 
as compared with a fermionic minicharges  model. In particular, we explain how the  sign of the ellipticity and rotation of the polarization plane acquired by a probe photon beam--in combination with their dependencies on the 
pulse parameters--can be exploited  to elucidate the quantum statistics of MCPs.

%%%%%%%%%%%%%%%%%%%%%%%%%%%%%%%%%%%%%%%%%%%%%%%%%%%%%%%%%%%%%%%%%%%%%%%%%%%%%%%%%%%%%%%%%%%%%%%%%%%%%%%%%%%%%%%%%%%%%%%%%%%%%%%%%
\section{Theoretical aspects}
%%%%%%%%%%%%%%%%%%%%%%%%%%%%%%%%%%%%%%%%%%%%%%%%%%%%%%%%%%%%%%%%%%%%%%%%%%%%%%%%%%%%%%%%%%%%%%%%%%%%%%%%%%%%%%%%%%%%%%%%%%%%%%%%%

%%%%%%%%%%%%%%%%%%%%%%%%%%%%%%%%%%%%%%%%%%%%%%%%%%%%%%%%%%%%%%%%%%%%%%%%%%%%%%%%%%%%%%%%%%%%%%%%%%%%%%%%%%%%%%%%%%%%%%%%%%%%%%%%%
\subsection{Photon Green's function and vacuum polarization \label{opkimixing}}
%%%%%%%%%%%%%%%%%%%%%%%%%%%%%%%%%%%%%%%%%%%%%%%%%%%%%%%%%%%%%%%%%%%%%%%%%%%%%%%%%%%%%%%%%%%%%%%%%%%%%%%%%%%%%%%%%%%%%%%%%%%%%%%%%

It is a long-standing prediction of QED that the optical properties of its vacuum are modified in the presence of an external electromagnetic 
field due to the nontrivial interplay between photons and the fluctuations of  virtual electron-positron pairs polarized by an external 
field.  Indeed, compelling theoretical studies  provide  evidences for self-coupling  of photons, rendering QED a nonlinear theory which allows 
for birefringence  and absorption of photons traveling through the  polarized region  of the  vacuum.  However, the source  of fluctuation  
inducing nonlinear self-interactions of the  electromagnetic  field is not restricted  to virtual electrons and positrons. Although at the energy 
scale of QED,  the  structure  of the quantum vacuum is mainly determined by these virtual entities, actually any quantum degree  of freedom  
that couples to photons  modifies the structure  of the effective  vertices  which result  from  the  generating functional of the one-particle 
irreducible Feynman graphs. The lowest one, i.e., the one  containing two amputated legs:\footnote{From now  on ``natural'' and Gaussian  units with  $c=\hbar=4\pi\epsilon_0=1$ are used.} 
\begin{eqnarray}
\mathscr{D}_{\mu\nu}^{-1}(k,k^\prime)=-\frac{1}{4\pi}\left(k^2\mathpzc{g}_{\mu\nu}-k_{\mu} k_{\nu}\right)\deltabar_{k,k^\prime}+\frac{1}{4\pi}\Pi_{\mu\nu}(k,k^\prime),\label{twopointfunction}
\end{eqnarray}defines the vacuum polarization tensor $\Pi_{\mu\nu}(k,k^\prime)$  through the Green's function of MCPs  as well as the bare and dressed vertices, as it occurs in a pure QED context. Here  $\mathpzc{g}_{\mu\nu}=\mathrm{diag}(+1,-1,-1,-1)$  denotes the flat metric tensor, whereas 
the shorthand notation $\deltabar_{k,k^\prime}=(2\pi)^4\delta^4(k-k^\prime)$ has been introduced. 

In the one-loop approximation,  and in the  field of a circularly polarized  monochromatic plane-wave of the form
\begin{eqnarray}
\begin{array}{c}
\displaystyle \mathscr{A}^{\mu}(x)=\mathpzc{a}_1^{\mu}\cos(\varkappa x)+\mathpzc{a}_2^{\mu}\sin(\varkappa x)\\
\varkappa=(\varkappa^0,\pmb{\varkappa}),\quad \varkappa \mathpzc{a}_i=0,\quad \varkappa^2=0\quad \mathrm{and}\quad \mathpzc{a}_1^2=\mathpzc{a}_2^2\equiv\mathpzc{a}^2
\end{array}
\label{externalF}
\end{eqnarray} the  polarization tensor  splits  into elastic and inelastic terms:
\begin{eqnarray}\label{ptcircular}
\begin{array}{c}\displaystyle
\Pi^{\mu\nu}(k,k^\prime)=\deltabar_{k,k^\prime}\Pi^{\mu\nu}_0(k^\prime)+\deltabar_{k,k^\prime-2\varkappa}\Pi^{\mu\nu}_-(k^\prime)+\deltabar_{k,k^\prime+2\varkappa}\Pi^{\mu\nu}_+(k^\prime),\\
\displaystyle\Pi^{\mu\nu}_0(k^\prime)=-\sum_{i=\pm,\parallel} \pi_i(k^\prime)\Lambda^\mu_i(k^\prime)\Lambda^{\nu*}_i(k^\prime),\quad \Pi^{\mu\nu}_\pm(k^\prime)=2\pi_0(k^\prime)\Lambda_\pm^\mu(k^\prime)\Lambda_\pm^\nu(k^\prime)
\end{array}
\end{eqnarray}out of which  the elastic  contribution   $\Pi^{\mu\nu}_0(k^\prime)$  is diagonalizable. Its eigenvalues $\pi_i$, as well as the form factor $\pi_0$, are functions which  
have been evaluated thoroughly for the case of spinor and scalar QED in \cite{baier}.  In constrast to  $\Pi^{\mu\nu}_0(k^\prime)$, the other two  terms in Eq.~(\ref{ptcircular}) describe  
inelastic  processes  characterized by  the   emission or  absorption of photons of the high-intensity laser wave.  The involved  eigenvectors $\Lambda_+(k^\prime),$ $\Lambda_-(k^\prime)$ and 
$\Lambda_\parallel(k^\prime)$ are transverse  $k^\prime\cdot\Lambda_j(k^\prime) =0$, orthogonal to each other--$\Lambda^\prime_{i}(k^\prime)\cdot\Lambda_{j}(k^\prime)=-\delta_{ij}$, and fulfill 
the completeness relation 
\begin{equation}
\mathpzc{g}^{\mu\nu}-\frac{k^{\prime\mu} k^{\prime\nu}}{k^{\prime2}}=-\sum_{i=\pm,\parallel}\Lambda^\mu_{i}(k^\prime)\Lambda^\nu_{i}(k^\prime). \label{completeness}
\end{equation}Particularly, we have that  $\Lambda_\pm$  turn out to be  eigenstates of opposite helicities with  $\Lambda^*_\pm=\Lambda_\mp$. 

In its simple version, a  scenario involving MCPs characterized by a mass $m_\epsilon$ and  a tiny fraction of the electron charge $\mathpzc{q}_\epsilon\equiv\epsilon\vert e\vert$ is reminiscent of QED;  
the  phenomenological consequences associated with their  existence  would  not differ qualitatively from  those emerging in a pure QED context. As such,  one can  investigate the related processes from already known 
QED expressions,  with  the electron  parameters $(e,\ m)$   replaced by the respective quantities associated with a MCP $(\mathpzc{q}_\epsilon,\ m_\epsilon)$. So, in the following,  we evaluate the extent 
to which  MCPs might influence  the  propagation of a probe  photon in the field of the strong laser wave  [Eq.~(\ref{externalF})] through the dispersion laws that  result  from the poles of  the   photon Green's 
function  $\mathscr{D}_{\mu\nu}(k,k^\prime)$. The latter can be obtained by  inversion of the two-point irreducible function [Eq.~(\ref{twopointfunction})],  since   
\[
\int\frac{d^4k^{\prime\prime}}{(2\pi)^4}\mathscr{D}_{\mu\rho}^{-1}(k,k^{\prime\prime})\mathscr{D}^{\rho\nu}(k^{\prime\prime},k^\prime)=-\delta^\nu_{\mu}\deltabar_{k,k^\prime}.
\] Indeed, by inserting the decomposition of the polarization tensor we find that--up to an inessential  longitudinal contribution--the photon Green function in the field of the wave [Eq.~(\ref{externalF})] is given by 
\begin{equation}
\begin{array}{c}
\displaystyle 
\mathscr{D}^{\mu\nu}(k,k^\prime)=\deltabar_{k,k^\prime}\sum_{i=\pm,\parallel} \mathscr{D}_{j}^{\mathrm{el}}\Lambda^\mu_j\Lambda^{\nu*}_j+\sum_{j=\pm}\deltabar_{k,k^\prime-2j\varkappa}\mathscr{D}_j^{\mathrm{in}}\Lambda_j^\mu\Lambda_j^\nu\\
\displaystyle\mathscr{D}_{\pm}^{\mathrm{el}}=-\frac{k_\pm^2-\pi_\mp(k_\pm)}{\left[k^2-\pi_\pm(k)\right]\left[k_\pm^2-\pi_\mp(k_\pm)\right]-4\pi_0(k)\pi_0(k_\pm)},\\ 
\displaystyle \mathscr{D}_{\parallel}^{\mathrm{el}}=-\frac{1}{k^2-\pi_\parallel(k)},\quad \mathscr{D}_{\pm}^{\mathrm{in}}=-\frac{2\pi_0(k_\pm)}{\left[k^2-\pi_\pm(k)\right]\left[k_\pm^2-\pi_\mp(k_\pm)\right]-4\pi_0(k)\pi_0(k_\pm)},
\end{array}
\label{greencircular}
\end{equation}where $k_\pm\equiv k\pm 2\varkappa$ and $k=(\mathpzc{w},\pmb{k})$. We remark that, in deriving the Green's function  the completeness relation [Eq.~(\ref{completeness})] has  been  taken  into account.

Hereafter we consider the limiting case in which the polarization effects  due to MCPs are  tiny  corrections to  the free photon dispersion equation [$k^2\simeq0$]. In  this approximation, 
the pole  associated with the $\parallel$-mode does not correspond to photon type excitations, since--independently of the $\pi_\parallel$-structure--the corresponding  eigenvector  $\Lambda_\parallel$ becomes 
purely longitudinal at $k^2=0$   [more details can be found in page 7 of part I of this series]. Conversely, the dispersion equations resulting from the poles associated with  the  transverse modes $\Lambda_\pm$  
concide with those found previously  in  Refs.~\cite{baier,Villalba-Chavez:2013gma,Villalba-Chavez:2014nya}:
\begin{equation}\label{dispersionequation}
\left(k^2-\pi_\pm(k)\right)\left[k_\pm^2-\pi_\mp(k_\pm)\right]=4\pi_0(k)\pi_0(k_\pm).
\end{equation} The  corresponding vacuum refractive  indices $\mathpzc{n}_\pm^2(\mathpzc{w},\pmb{k}) =\pmb{k}^2/\mathpzc{w}^2=1-k^2/\mathpzc{w}^2$ turn out to be  
\begin{equation}\label{generaln}
\mathpzc{n}_\pm^2(\mathpzc{w},\pmb{k}) =1-\frac{\pi_\pm(k)}{\mathpzc{w}^2}-\frac{4\pi_0(k)\pi_0(k\pm2\varkappa)}{\mathpzc{w}^2[(k\pm2\varkappa)^2-\pi_\mp(k\pm2\varkappa)]}.
\end{equation} The last term in the right-hand side  of  Eq.~(\ref{generaln}) is responsible  for  inelastic transitions  between states with different  helicities.  In  the limit of interest  [$k^2\simeq0$]  
this formula reduces to
\begin{equation}
\mathpzc{n}_\pm(\pmb{k}) \simeq1-\frac{\pi_\pm(k)}{2\omega_{\pmb{k}}^2}\mp\frac{\pi_0(k)\pi_0(k_\pm)}{2\omega_{\pmb{k}}^2 k\varkappa},\label{intermediaterefrac}
\end{equation} where $\omega_{\pmb{k}}\equiv\vert\pmb{k}\vert$ denotes  the  energy of the probe photons. Hereafter, we restrict $\mathpzc{n}_\pm(\pmb{k})$ to an accuracy up to terms $\sim \pi_\pm/\omega_{\pmb{k}}^2$ 
so that the effects resulting from the last contribution  in Eq.~(\ref{intermediaterefrac}) are no longer considered. Note that this approximation is valid  as long as the condition
\begin{equation}
\frac{\pi_0(k_\pm)}{\omega_{\pmb{k}} \varkappa_0}\frac{\pi_0(k)}{\pi_\pm(k)}\ll1-\cos(\theta)
\end{equation} is satisfied; otherwise the use of our perturbative treatment would not be justified. We remark that, in this expression,  $\theta$  denotes the collision 
angle between the probe and the strong laser wave. For the particular situation to be studied later on, i.e.,  counterpropagating geometry  [$\theta=\pi$] with 
$\omega_{\pmb{k}}\sim \varkappa_0\sim1\  \rm eV$, the above condition would imply that  $\pi_0(k_\pm)\pi_0(k)/\pi_\pm(k)\ll 2\ \rm eV^2$  which can be   satisfied easily since 
the left-hand side is proportional to the square of the--presumably very tiny--coupling constant $\sim \epsilon^2 e^2$. Besides,  we will deal  with laser waves  
whose intensity parameters $\xi^2=-e^2\mathpzc{a}^2/m^2$ [with $m$ and $e$ the electron mass and charge, respectively] are smaller than unity. 

%%%%%%%%%%%%%%%%%%%%%%%%%%%%%%%%%%%%%%%%%%%%%%%%%%%%%%%%%%%%%%%%%%%%%%%%%%%%%%%%%%%%%%%%%%%%%%%%%%%%%%%%%%%%%%%%%%%%%%%%%%%%%%%%%%%%%%%%%%%%%%%%%%%%%%%%%%%%%
\subsection{Optical observables: including the  paraphoton interplay}
%%%%%%%%%%%%%%%%%%%%%%%%%%%%%%%%%%%%%%%%%%%%%%%%%%%%%%%%%%%%%%%%%%%%%%%%%%%%%%%%%%%%%%%%%%%%%%%%%%%%%%%%%%%%%%%%%%%%%%%%%%%%%%%%%%%%%%%%%%%%%%%%%%%%%%%%%%%%%

The  $\Pi^{\mu\nu}_0$-eigenvalues contain   real and imaginary contributions $\pi_\pm=\mathrm{Re}\ \pi_\pm+i\ \mathrm{Im}\ \pi_\pm$. The respective refractive indices--Eq.~(\ref{intermediaterefrac}) limited to the  
first two terms in the right-hand side--must also be  complex quantities, i.e., $\mathpzc{n}_\pm=n_\pm+i\varphi_\pm$. While the real part $n_\pm$ describes the pure  dispersive phenomenon, the imaginary
contribution  provides the absorption coefficient $\kappa_\pm=\varphi_\pm \omega_{\pmb{k}}$ for  mode-$\pm$ photons. Accordingly, we find in the limit under consideration that 
\begin{eqnarray}
n_\pm=1-\left.\frac{\mathrm{Re}\ \pi_\pm}{2 \omega_{\pmb{k}}^2}\right\vert_{k^2=0}\qquad \mathrm{and}\qquad \kappa_\pm=-\left.\frac{\mathrm{Im}\ \pi_\pm}{2\omega_{\pmb{k}}}\right\vert_{k^2=0}.
\label{refractioninde}
\end{eqnarray} Since the  analytic properties of $\mathrm{Re}\ \pi_+$ and $\mathrm{Re}\ \pi_-$ are different, the vacuum  behaves  like a chiral  
birefringent  medium.  As a consequence of this circumstance, the polarization  plane  of an incoming linearly polarized probe beam  rotates by a tiny angle:
\begin{eqnarray}\label{rot&elip}
\vert\vartheta(\epsilon,m_\epsilon)\vert \approx\frac{1}{2}\left\vert(n_+-n_-)\omega_{\pmb{k}}\tau\right\vert\ll1,
\end{eqnarray}where  $\tau$ is  the temporal pulse length. Besides, in the field of the laser wave the vacuum is predicted to be  dichroic. This  effect  induces a tiny ellipticity $\psi(\epsilon,m_\epsilon)$ in the 
polarization of the probe beam  which is determined by the nontrivial difference between the absorption coefficients
\begin{equation}\label{pureMCPellip}
\vert\psi(\epsilon,m_\epsilon)\vert \approx\frac{1}{2}\left\vert(\kappa_+-\kappa_-)\tau\right\vert\ll1\qquad .
\end{equation} The difference between  $\kappa_+$ and $\kappa_-$  manifests  by itself that  the photo-production rate of a  MCPs pair associated with  a  $\Pi_0^{\mu\nu}$-eigenwave  differs from the rate resulting from  the  
remaining mode. This statement is somewhat expected because the  optical theorem dictates that the  creation rate  of  a  pair from a probe photon with polarization vector $\Lambda_\pm$  is given by   
$\mathpzc{R}_{\ \pm}=\Lambda^{*\mu}_\pm\Lambda_\pm^{\nu}\rm Im\ \Pi_{0\mu\nu}/\omega=2\kappa_\pm$. We recall that the  energy-momentum  balance of this process $k+n\varkappa\to \mathpzc{q}_\epsilon^++ \mathpzc{q}_\epsilon^-$  allows us to  establish  the  threshold  condition $n\geqslant n_*$, where  $n_*=2m_\epsilon^2(1+\xi_\epsilon^2)/(k\varkappa)$  depends on the parameter  
$\xi_\epsilon^2=-\epsilon^2 e^2\mathpzc{a}^2/m_\epsilon^2$. In term of the MCP  mass $m_\epsilon$, the previous relation translates into  $m_\epsilon\leqslant m_n$,  with $m_n$ refering to the threshold mass
\begin{equation}
m_n\equiv\sqrt{\frac{1}{2} nk\varkappa-\epsilon^2m^2\xi^2}.\label{thresholdmass}
\end{equation} 

\begin{figure}
\begin{center}
\includegraphics[width=2.5in]{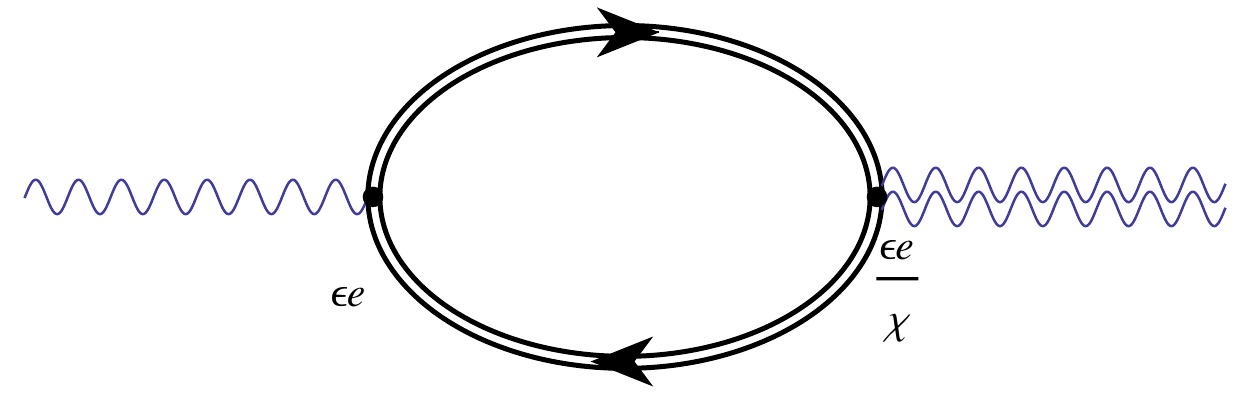}
\caption{\label{fig:mb001}Pictorial representation of the one-loop photon-paraphoton vertex.  The double lines represent the propagator of  MCPs including the full interaction with the external field. 
A  single wavy line denotes the amputated leg  corresponding to a  small-amplitude electromagnetic wave. Conversely,  a double wavy line refers to the  amputated leg associated with a  hidden-photon field.}
\end{center}
\end{figure}

The model described so far relies on a hypothetical existence of MCPs only. Their occurrence is nevertheless  naturally realized in scenarios involving hidden sectors containing  an extra $\rm U(1)$ gauge 
group. The corresponding hidden-photon field $w_\mu(x)$ is  massive with mass $m_{\gamma^\prime}$ and couples to the visible electromagnetic sector via a kinetic mixing characterized by an unknown parameter 
$\chi$. The diagonalization of this mixing term  induces  an effective  interaction between the hidden-current $j_h^\mu(x)$ and the total electromagnetic field $a_\mu(x)+\mathscr{A}(x)$: 
\begin{equation}\label{interactinglagra}
\mathpzc{L}_{\mathrm{int}}=-\chi e_h j_h^\mu(x)\left\{a_\mu(x)+\mathscr{A}_\mu(x)\right\}, 
\end{equation}where  $e_h$ refers to the hidden gauge coupling. In addition, a mass $m_\gamma=\chi m_{\gamma^\prime}$  for the visible electromagnetic field  $a_\mu(x)$  results. Furthermore, as a consequence 
of Eq.~(\ref{interactinglagra}), the  relation $\epsilon e=- \chi e_h$ is established and the two-point irreducible function  in the one-loop approximation becomes 
\begin{eqnarray}
\pmb{\mathpzc{D}}^{-1}(k,k^\prime)=-\frac{1}{4\pi}\left[
\begin{array}{cc}
\displaystyle (k^2-\chi^2m_{\gamma^\prime}^2)\mathpzc{g}_{\mu\nu}\deltabar_{k,k^\prime}-\Pi_{\mu\nu}(k,k^\prime)&\chi m_{\gamma^\prime}^2 \mathpzc{g}_{\mu\nu}\deltabar_{k,k^\prime}+\frac{1}{\chi}\Pi_{\mu\nu}(k,k^\prime)\\ \\
\chi m_{\gamma^\prime}^2\mathpzc{g}_{\mu\nu}\deltabar_{k,k^\prime}+\frac{1}{\chi}\Pi_{\mu\nu}(k,k^\prime)&  (k^2-m_{\gamma^\prime}^2)\mathpzc{g}_{\mu\nu}\deltabar_{k,k^\prime}-\frac{1}{\chi^2}\Pi_{\mu\nu}(k,k^\prime) 
\end{array}
\right].\nonumber
\end{eqnarray}Theoretical studies, as well as the experimental evidence indicate that the mixing parameter is much  smaller than unity [$\chi\ll1$] so that a perturbative treatment in $\chi$ is well suited. With 
such an approximation, the mass term of the electromagnetic field can be ignored,  leading to  describe the probe photon beam by two transverse polarization states  $\Lambda_\pm$, whereas the $\Lambda_\parallel-$mode 
remains longitudinal and  unphysical.\footnote{This fact contrasts with the phenomenology  occurring in a plasma, where the  polarization tensor provides a longitudinal mode due to the nontrivial interplay  with 
the medium. Indeed, if the  plasma is nonrelativistic, a photon behaves like a particle with a mass determined by the plasma frequency. Under such a condition, the oscillation rate between longitudinal modes is predicted 
to be  larger  than the corresponding one between transversal modes. The relevance of this fact in deriving  constraints from solar  luminosity arguments has been put foward in Ref.~\cite{An:2013yfc}.}

Observe that the off-diagonal terms in $\pmb{\mathpzc{D}}^{-1}(k,k^\prime)$  allow for  the photo-paraphoton oscillation, a  process driven by both:  the  massive terms  
$\chi m_{\gamma^\prime}^2\mathpzc{g}_{\mu\nu}\deltabar_{k,k^\prime}$ and those involving the  vacuum polarization tensor $\frac{1}{\chi}\Pi_{\mu\nu}(k,k^\prime)$. However, hereafter we will suppose  
that the energy scale provided by the loop is much greater than  the scale associated with  the paraphoton mass [$\chi^2 m_\gamma^2\ll \pi_\pm$] which only leaves room for oscillations mediated by 
virtual pairs of MCPs [see Fig.~\ref{fig:mb001}].  As a consequence of this hypothetical  phenomena,  the  polarization plane of a linearly-polarized  probe beam  should be rotated by an angle
\begin{eqnarray}\label{rotation}
&&\vert\vartheta(\epsilon,m_\epsilon,\chi)\vert \approx\frac{1}{2}\left\vert(n_+-n_-)\omega_{\pmb{k}}\tau+\chi^2\sin\left(\frac{n_+-1}{\chi^2}\omega_{\pmb{k}}\tau\right)\exp\left(-\frac{1}{\chi^2}\kappa_+ \tau\right)\right.\nonumber\\
&&\qquad\qquad\qquad\qquad\qquad\qquad-\left.\chi^2\sin\left(\frac{n_--1}{\chi^2}\omega_{\pmb{k}}\tau\right)\exp\left(-\frac{1}{\chi^2}\kappa_- \tau\right)\right\vert\ll1.
\end{eqnarray} Observe that the first contribution coincides with the outcome resulting from a pure MCPs model [Eq.~(\ref{rot&elip})]. Hence, those  terms that depend on the unknown parameter 
$\chi$ are connected to the photon-paraphoton  oscillations.  

The scenario including the hidden-photon field manifests vacuum dichroism as well, since the decay rates for  the two  ``visible'' $\Pi_{0}^{\mu\nu}$-eigenmodes,  via the production 
of a MCPs pairs and  its conversion into a hidden-photon, differ from each other. The predicted  ellipticity  is   determined by the difference between the attenuation coefficients of the propagating modes. 
Explicitly,
\begin{eqnarray}\label{ellipticity}
&&\vert\psi(\epsilon,m_\epsilon,\chi)\vert \approx\frac{1}{2}\left\vert(\kappa_--\kappa_+)\tau+\chi^2\cos\left(\frac{n_+-1}{\chi^2}\omega_{\pmb{k}}\tau\right)\exp\left(-\frac{1}{\chi^2}\kappa_+ \tau\right)\right.\nonumber\\
&&\qquad\qquad\qquad\qquad\qquad\qquad-\left.\chi^2\cos\left(\frac{n_--1}{\chi^2}\omega_{\pmb{k}}\tau\right)\exp\left(-\frac{1}{\chi^2}\kappa_- \tau\right)\right\vert\ll1.
\end{eqnarray}Note that in the absence of the kinetic mixing [$\chi\to 0$] this expression  reduces to Eq.~(\ref{pureMCPellip}). Throughout our investigation,  comparisons  between the  pure MCPs model  and the  scenario including 
the paraphotons will be presented.

%%%%%%%%%%%%%%%%%%%%%%%%%%%%%%%%%%%%%%%%%%%%%%%%%%%%%%%%%%%%%%%%%%%%%%%%%%%%%%%%%%%%%%%%%%%%%%%%%%%%%%%%%%%%%%%%%%%%%%%%%%%%%%%%%
\subsection{Absorption coefficients and refractive indices at  $\xi_\epsilon<1$ \label{spectraldecompos}}
%%%%%%%%%%%%%%%%%%%%%%%%%%%%%%%%%%%%%%%%%%%%%%%%%%%%%%%%%%%%%%%%%%%%%%%%%%%%%%%%%%%%%%%%%%%%%%%%%%%%%%%%%%%%%%%%%%%%%%%%%%%%%%%%%

In contrast to part I of this series, here  we analyse the  effects resulting from a model in which MCPs are scalar bosons. In the first place, the absence of a spin in these hypothetical 
degrees of  freedom is manifest in the eigenvalues:
\begin{equation}\label{formfactors}
\pi_\pm(n_*,\xi_\epsilon)=\frac{\alpha_\epsilon}{2\pi} m_\epsilon^2\int_{-1}^{1}dv \int_0^\infty \frac{d\rho}{\rho}\Omega_\pm \exp\left\{-\frac{2i\rho n_*}{(1+\xi_\epsilon^2)(1-v^2)}\left[1+2A\xi_\epsilon^2\right]\right\}.
\end{equation} In this expression,  $\alpha_\epsilon\equiv \epsilon^2 e^2=\epsilon^2/137$ refers to the fine structure constant relative to the MCPs [with $\epsilon$ being   
the potentially small coupling strength in units of the absolute value of the electron charge $\vert e\vert$].  The expression above depends on the threshold parameter for the photo-production of a  pair of MCPs $n_*=2m_\epsilon^2(1+\xi_\epsilon^2)/(k\varkappa)$ 
[see discussion  above Eq.~(\ref{thresholdmass})]. Here, the functions $\Omega_\pm$ and $A$ read 
\begin{eqnarray}
\begin{array}{c}\displaystyle 
\Omega_\pm=\xi_\epsilon^2\left[\sin^2(\rho)\pm2i\rho A_0\right]+\frac{1}{2}\left[1-\exp(iy)\right],\quad A=\frac{1}{2}\left[1-\frac{\sin^2(\rho)}{\rho^2}\right],\\  
\displaystyle
 A_0=\frac{1}{2}\left[\frac{\sin^2(\rho)}{\rho^2}-\frac{\sin(2\rho)}{2\rho}\right],\quad y=\frac{4 n_* \xi_\epsilon^2 \rho A}{(1+\xi_\epsilon^2)(1-v^2)}.
\end{array}
\label{g3fermion}
\end{eqnarray}When integrating out $v$,   a compact representation of Eq.~(\ref{formfactors}) is  obtained.  However, the resulting integrand involves  unwieldy complex functions depending 
on the Hankel functions of second kind $\mathrm{H}_{\nu}^{(2)}(z)=\frac{2i}{\pi}\exp[\frac{i}{2}\pi\nu]\int_0^{\infty} dt \exp[-iz\cosh(t)]\cosh(\nu)$. Explicitly, we find
\begin{eqnarray}
\label{eigenvalues}
&&\pi_\pm(n_*,\xi_\epsilon)=\frac{1}{2} \alpha_\epsilon m_\epsilon^2 \int_0^\infty \frac{d\rho}{\rho}\Upsilon_\pm \exp\left(-i\eta\right),\\
&&\Upsilon_\pm=-\frac{1}{2}\eta\left[\mathrm{H}_0^{(2)}(\eta)+i\mathrm{H}_1^{(2)}(\eta)\right]\left\{1+2\xi_\epsilon^2\sin^2(\rho)\pm 4i\xi_\epsilon^2 \rho A_0\right\}\nonumber\\ 
&&\qquad\qquad\qquad\qquad\qquad\qquad+\frac{1}{2}\rho n_* \left[\mathrm{H}_0^{(2)}(\rho n_*)+i\mathrm{H}_1^{(2)}(\rho n_*)\right]\exp\left[\frac{2i\xi_\epsilon^2 \rho n_* A }{1+\xi_\epsilon^2}\right],\label{scalarintegrandsei}
\end{eqnarray}where the parameter $\eta\equiv n_*\rho(1-\Delta)$ with $\Delta=\frac{\xi_\epsilon^2}{(1+\xi_\epsilon^2)}\frac{\sin^2(\rho)}{\rho^2}$ has been introduced.

As in I, our attention will be focused on the limit  $\xi_\epsilon<1$. Particularly, on the simple cases in which one or two  photons from the strong wave [$n_*=1,2$] are absorbed.  We will consider 
these two situations  only  because--for  $\xi_\epsilon<1$--the  chiral  birefringence and dichroism  properties of the vacuum are predicted to be considerably more  pronounced near  the lowest 
thresholds  than in the cases asymptotically far from it [$n_*\to \infty$ and $n_*\to0$], where  the vacuum behaves like a nonabsorbing isotropic medium  
\cite{Villalba-Chavez:2013gma}.   %So, contributions of higher thresholds  $[n>2]$ are beyond the scope of this work. 
Note that in the region of interest [$\xi_\epsilon<1$], the parameter $\Delta$  is much   smaller  than  unity. So,  we may Taylor expand the integrands in Eqs.~(\ref{eigenvalues})-(\ref{scalarintegrandsei}) 
up to second order in $\Delta$ and  integrate out the $\rho-$variable. The real  parts of the resulting expressions allows us to write the absorption coefficients [Eq.~(\ref{refractioninde})] in the following form: 
\begin{equation}\label{absorptioncoeffdecompo}
\kappa_\pm\simeq\kappa_{\pm,1}+\kappa_{\pm,2}. 
\end{equation} Here  $\kappa_{\pm,1}$  and $\kappa_{\pm,2}$ turn out to be discontinuous contributions at the threshold point $n_*=1$ and $n_*=2$, respectively. Particularly, we find 
\begin{eqnarray}
\label{scalarpositivekappa}
&&\kappa_{+,1}=\frac{\alpha_\epsilon m^2_\epsilon \xi_\epsilon^2}{8\omega_{\pmb{k}}}\left\{\mathpzc{v}_1\frac{1-\mathpzc{v}_1^2}{1+\xi_\epsilon^2}+\left[1-\mathpzc{v}_1^2-\frac{1-\mathpzc{v}_1^4}{2(1+\xi_\epsilon^2)}\right]\ln\left(\frac{1+\mathpzc{v}_1}{1-\mathpzc{v}_1}\right)\right\}\Theta[\mathpzc{v}_1^2],\\  
&&\kappa_{-,1}=\frac{\alpha_\epsilon m^2_\epsilon \xi_\epsilon^2}{8\omega_{\pmb{k}}}\left\{\mathpzc{v}_1\left(2+\frac{1-\mathpzc{v}_1^2}{1+\xi_\epsilon^2}\right)-\left[1-\mathpzc{v}_1^2+\frac{1-\mathpzc{v}_1^4}{2(1+\xi_\epsilon^2)}\right]\ln\left(\frac{1+\mathpzc{v}_1}{1-\mathpzc{v}_1}\right)\right\}\Theta[\mathpzc{v}_1^2],\label{scalarnegativekappa}
%&&\kappa_{+,1}=\frac{\alpha_\epsilon m^2_\epsilon \xi_\epsilon^2}{4\omega_{\pmb{k}}}\left\{\frac{1-\mathpzc{v}_1^4}{2(1+\xi_\epsilon^2)}\ln\left(\frac{1+\mathpzc{v}_1}{1-\mathpzc{v}_1}\right)+2\mathpzc{v}_1\left(1-\frac{1-\mathpzc{v}_1^2}{2(1+\xi_\epsilon^2)}\right)\right\}\Theta[\mathpzc{v}_1^2],\label{fermionpositivekappa} \\ 
%&&\kappa_{-,1}=\frac{\alpha_\epsilon m^2_\epsilon \xi_\epsilon^2}{4\omega_{\pmb{k}}}\left\{\left(2+\frac{1-\mathpzc{v}_1^4}{2(1+\xi_\epsilon^2)}\right)\ln\left(\frac{1+\mathpzc{v}_1}{1-\mathpzc{v}_1}\right)-4\mathpzc{v}_1\left(1+\frac{1-\mathpzc{v}_1^2}{4(1+\xi_\epsilon^2)}\right)\right\}\Theta[\mathpzc{v}_1^2].\label{fermionnegativekappa}
\end{eqnarray} where  $\Theta[x]$ denotes  the unit step  function and  $\mathpzc{v_1}=(1-n_*)^{\nicefrac{1}{2}}$ determines the relative speed between the  final particles $\vert\pmb{\mathpzc{v}}_{\mathrm{rel}}\vert=2\mathpzc{v}_1$ 
when  only one   photon of the intense  laser wave is  absorbed.  We emphasize that  Eqs.~(\ref{scalarpositivekappa})-(\ref{scalarnegativekappa})  provide  nonvanishing 
contributions whenever the MCP mass $m_\epsilon$ is smaller or equal to the first  threshold mass  $m_1=\left(k\varkappa/2-\epsilon^2m^2\xi^2\right)^{\nicefrac{1}{2}}$, 
corresponding to $n_*\leqslant1$.  Conversely, the contributions resulting from  the absorption of two photons of the laser wave is valid for masses  $m_\epsilon<m_2=\left(k\varkappa-\epsilon^2m^2\xi^2\right)^{\nicefrac{1}{2}}$. They amount to
\begin{eqnarray}\label{absorptioncoefficient2order}
\kappa_{\pm,2}=\frac{\alpha_\epsilon m^2_\epsilon \xi_\epsilon^4}{4\omega_{\pmb{k}}(1+\xi_\epsilon^2)}\left[\mathpzc{F}_1(\mathpzc{v}_2)+2\frac{1-\mathpzc{v}_2^2}{1+\xi_\epsilon^2}\mathpzc{F}_2(\mathpzc{v}_2)\pm \mathpzc{F}_3(\mathpzc{v}_2)\right]\Theta[\mathpzc{v}_2^2],
\end{eqnarray}where $\mathpzc{v}_2=(1-n_*/2)^{\nicefrac{1}{2}}$ and  the functions $\mathpzc{F}_i(\mathpzc{v}_2)$  with $i=1,2,3$ are given by 
\begin{eqnarray}\label{F1fermion}
&&\mathpzc{F}_1(\mathpzc{v}_2)=\mathpzc{v}_2\left(1+\mathpzc{v}_2^2\right)-\left(1-\mathpzc{v}_2^2\right)^2\arctanh(\mathpzc{v}_2),\\ 
&&\mathpzc{F}_2(\mathpzc{v}_2)=\frac{1}{12}\mathpzc{v}_2\left(15\mathpzc{v}_2^4-4\mathpzc{v}_2^2-3\right)+\frac{1}{4}\left(1+\mathpzc{v}_2^2+3\mathpzc{v}_2^4-5\mathpzc{v}_2^6\right)\arctanh(\mathpzc{v}_2),\\
&&\mathpzc{F}_3(\mathpzc{v}_2)=\frac{1}{4}\mathpzc{v}_2\left(1+3\mathpzc{v}_2^2\right)-\frac{1}{4}\left(1+3\mathpzc{v}_2^4\right)\arctanh(\mathpzc{v}_2).\label{F3scalar}
\end{eqnarray}Some comments are in order. Firstly, Eqs.(\ref{absorptioncoefficient2order})-(\ref{F3scalar}) were determinated by restricting the threshold parameter  to  $1< n_*\leq2$, so that the next-to-leading order contribution  
[$\sim\xi^4_\epsilon$] to the  two-photon  reaction is not considered. We remark that,  when the scalar MCPs  are created in the center-of-mass frame almost at rest [$\mathpzc{v}_2\sim0$ corresponding 
to $n_*\to 2$],  the functions $\mathpzc{F}_i(\mathpzc{v}_2)$ are dominated by the cubic dependences on $\mathpzc{v}_2$. As a consequence, the absorption coefficients for the  scalar theory  approach to  $\kappa_{\pm,2}\approx\alpha_\epsilon m^2_\epsilon \xi_\epsilon^4 \mathpzc{v}_2^3(8\mp1)/[12\omega_{\pmb{k}}(1+\xi_\epsilon^2)]$. 
Conversely, when $n_*\to1$, i.e., [$\mathpzc{v}_2\to\nicefrac{1}{\sqrt{2}}$],  we find  the asymptotes   $\kappa_{\pm,2}\approx\alpha_\epsilon m^2_\epsilon \xi_\epsilon^4 (0.4\mp0.1)/[4\omega_{\pmb{k}}]$, provided 
the condition $\xi_\epsilon\ll1$ holds. The corresponding expression for $\kappa_{\pm,1}$ was derived  previously in Ref.~\cite{Villalba-Chavez:2013gma}.

In contrast to $\mathrm{Re}\ \pi_\pm$,  the imaginary parts of  $\pi_\pm$  are continuous functions. Hence, we only need to consider the refractive indices [Eq.~(\ref{refractioninde})] resulting from the 
leading term  in the  $\Delta$-expansion of the respective integrands [see Eqs.~(\ref{eigenvalues})-(\ref{scalarintegrandsei})] in order to investigate the  dispersion effects in the  region  encompassed  by 
Eqs.~(\ref{absorptioncoeffdecompo})-(\ref{F3scalar}), i.e., $0<n_*\leq2$. After some manipulations, we end up with an integral representation for $n_\pm-1$,  suitable  to carry out the forthcoming numerical analysis
\begin{eqnarray}
&&n_\pm-1\simeq\mp\frac{\alpha_\epsilon m_\epsilon^2 \xi_\epsilon^2}{4\pi\omega_{\pmb{k}}^2}\int_{0}^1dv\left\{\left[1\pm \frac{2\varrho}{1+\xi_\epsilon^2}\right]\ln\left(\frac{1+\varrho}{\vert1-\varrho\vert}\right)^{\nicefrac{1}{2}}\right.\nonumber\\ 
&&\qquad\qquad\qquad\qquad\qquad\qquad\qquad\quad\left.\mp\left[1\pm2\varrho+\frac{2\varrho^2}{1+\xi_\epsilon^2}\right]\ln\left(\frac{\vert\varrho\vert}{\sqrt{\vert1-\varrho^2\vert}}\right)\right\}.\label{interpi0}
%&& n_\pm-1\simeq\pm\frac{\alpha_\epsilon m_\epsilon^2  \xi_\epsilon^2}{2\pi\omega_{\pmb{k}}^2} \int_{0}^1dv\left\{\left[1-\frac{2\varrho}{n_*}\left(1\mp \frac{n_*}{1+\xi_\epsilon^2}\right)\right]\ln\left(\frac{1+\varrho}{\vert1-\varrho\vert}\right)^{\nicefrac{1}{2}}\right.\nonumber\\ 
%&&\qquad\qquad\left.\mp\left[1-\frac{2\varrho}{n_*}(1\mp n_*)+\frac{2\varrho^2}{1+\xi_{\epsilon}^2}\left(1\mp\frac{2(1+\xi_\epsilon^2)}{n_*}\right)\right]\ln\left(\frac{\vert\varrho\vert}{\sqrt{\vert1-\varrho^2\vert}}\right)\right\}\label{interpi3},
\end{eqnarray} In this expression, $\varrho\equiv\varrho(v,n_*)=n_*(1-v^2)^{-1}$ is a function of both the integration variable $v$ and the threshold parameter $n_*$.   
 
%%%%%%%%%%%%%%%%%%%%%%%%%%%%%%%%%%%%%%%%%%%%%%%%%%%%%%%%%%%%%%%%%%%%%%%%%%%%%%%%%%%%%%%%%%%%%%%%%%%%%%%%%%%%%%%%%%%%%%%%%%%%%%%%%
\section{Experimental prospects   \label{PES}}
%%%%%%%%%%%%%%%%%%%%%%%%%%%%%%%%%%%%%%%%%%%%%%%%%%%%%%%%%%%%%%%%%%%%%%%%%%%%%%%%%%%%%%%%%%%%%%%%%%%%%%%%%%%%%%%%%%%%%%%%%%%%%%%%%

%%%%%%%%%%%%%%%%%%%%%%%%%%%%%%%%%%%%%%%%%%%%%%%%%%%%%%%%%%%%%%%%%%%%%%%%%%%%%%%%%%%%%%%%%%%%%%%%%%%%%%%%%%%%%%%%%%%%%%%%%%%%%%%%%
\subsection{Estimating the exclusion limits   \label{PAPH}}
%%%%%%%%%%%%%%%%%%%%%%%%%%%%%%%%%%%%%%%%%%%%%%%%%%%%%%%%%%%%%%%%%%%%%%%%%%%%%%%%%%%%%%%%%%%%%%%%%%%%%%%%%%%%%%%%%%%%%%%%%%%%%%%%%

Let us  estimate  the projected bounds resulting from a plausible experiment in which the  rotation of the polarization plane [Eqs.~(\ref{rotation})] and  the ellipticity 
of the outgoing probe beam  [Eq.~(\ref{ellipticity})] are probed  but none of them  detected. In practice, the absence of these signals is  understood  within  certain confidence levels $\psi_{\mathrm{CL}\%}$,  
$\vartheta_{\mathrm{CL}\%}$, which we take herafter as  $\sim10^{-10}\ \mathrm{rad}$. We emphasize that this choice of sensitivity  agrees with the experimental accuracies  with 
which--in  the optical regime--both observables can nowadays be  measured \cite{Muroo_2003}. Thus, in the following we present the numeric outcomes resulting from the inequalities  
\begin{equation}
10^{-10}\ \mathrm{rad}>\vert\psi(\epsilon,m_\epsilon,\chi)\vert\quad \mathrm{and}\quad 10^{-10}\ \mathrm{rad}>\vert\vartheta(\epsilon,m_\epsilon,\chi)\vert. 
\end{equation} Some comments are in order. Firstly, the  sensitivity limits found from these relations  will be  close to reality as the parameters of the external field [Eq.~(\ref{externalF})] will be chosen  
 appropriately  to the monochromatic plane-wave model. In an actual experimental setup this restriction can be met by using a long pulse of duration $\tau\gg \varkappa_0^{-1}$ whose waist size $w_0$ 
is much greater than its  wavelength  [$w_0\gg\lambda_0$ with $\lambda_0=2\pi\varkappa_0^{-1}$]. In this way, a  negligible contribution  coming from the finite bandwidth is  guaranteed. Based on 
the previous remarks, we it find suitable to  consider the benchmark parameters associated with the nanosecond frontend of the Petawatt  High-Energy Laser for heavy Ion eXperiments (PHELIX) \cite{PHELIX}, 
[$\tau\simeq20 \ \rm ns$, $w_0\approx 100-150 \mu \mathrm{m}$, $\varkappa_0\simeq1.17\ \rm eV$, $I\simeq 10^{16}\  \rm W/cm^2$, $\xi\simeq 6.4\times 10^{-2}$]. We also investigate the results coming from 
the parameters associated with the nanosecond facility of the  LULI(2000) system  \cite{LULI}, [$\tau\simeq 1.5-4 \ \rm ns$, $w_0\sim100 \ \mu\mathrm{m}$, $\varkappa_0\simeq1.17\ \rm eV$, $I\simeq6\times 10^{14}\ \rm W/cm^2$, 
$\xi\simeq2\times 10^{-2}$]. Clearly, with this  second analysis  we seek  to evaluate the extent to which the projected bounds depend on the parameters of the external field. 

Now, a suitable experimental development requires a high level of synchronization between the colliding laser waves. To guarantee this important aspect, it appears convenient to use a probe obtained from 
the intense  wave. So, we will assume a probe beam  with doubled frequency [$\omega_{\pmb{k}}=2\varkappa_0=2.34\ \rm eV$]   and  an  intensity  much smaller than the one of  the strong laser field. 
Finally, to maximize the polarimetric effects, we will suppose that the collision between the probe and strong wave  is head-on [$\pmb{k}\cdot\pmb{\varkappa}=-\omega_{\pmb{k}}\varkappa_0$].

\begin{figure}
\includegraphics[width=.48\textwidth]{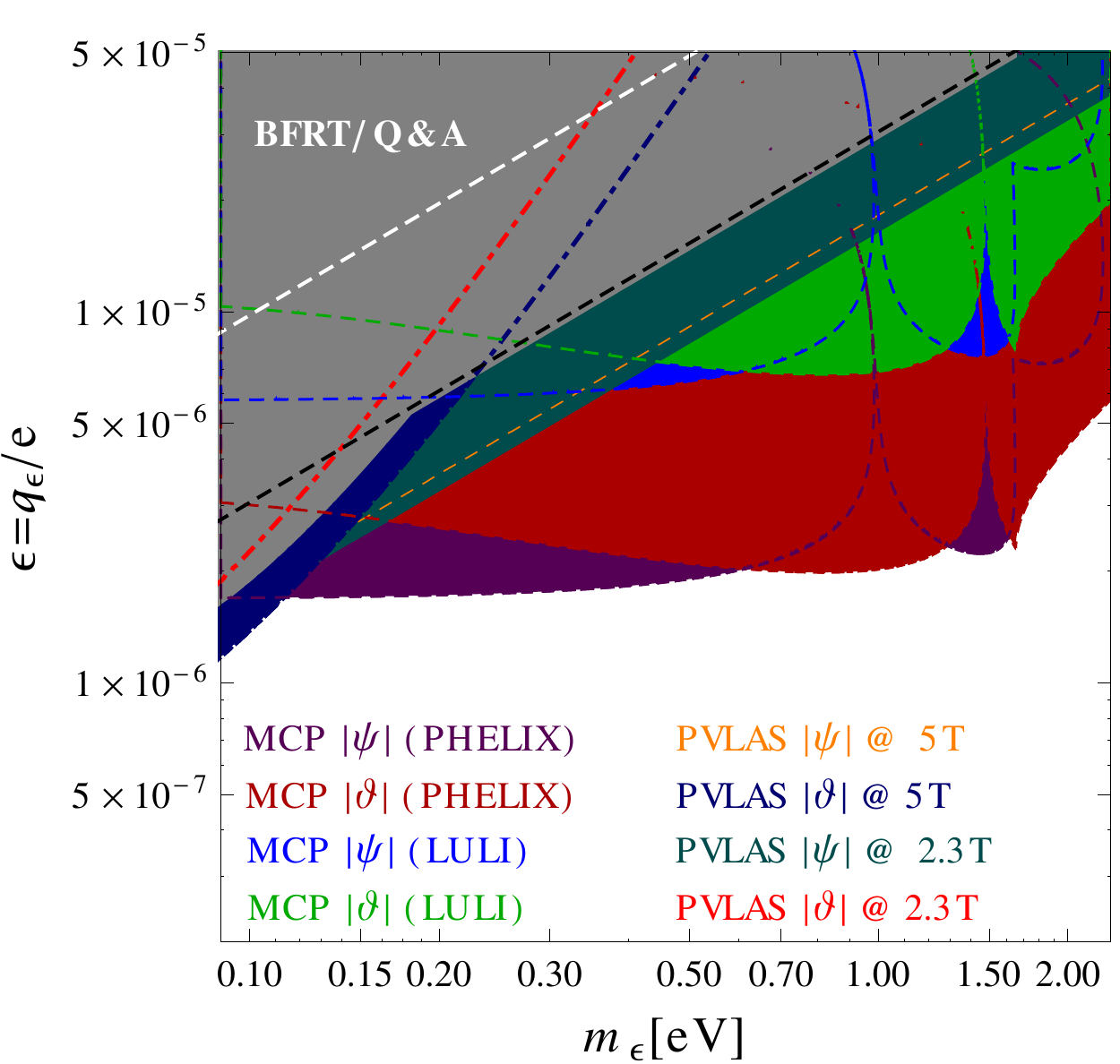}
\hspace{0.1cm}
\includegraphics[width=.48\textwidth]{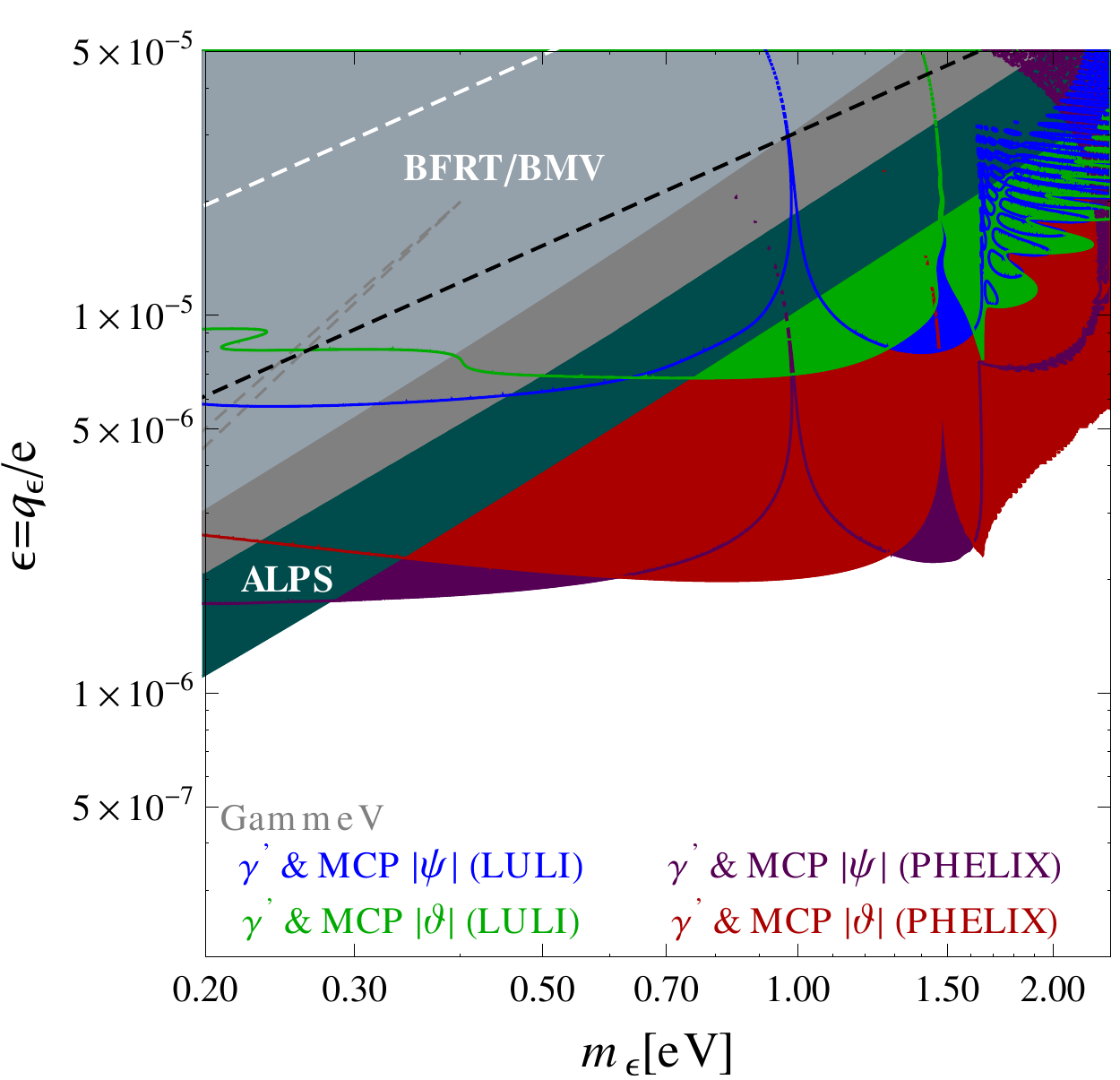}
\caption{\label{fig:mb003}Estimates of  constraints for MCPs  of  mass $m_\epsilon$ and relative coupling constant $\epsilon$ derived  from the absence of signals in  
a plausible polarimetric setup assisted by a circularly polarized  laser field of moderate intensity. While the left panel provides the results associated  purely with MCPs, 
the right one shows the outcomes of the  model including a hidden-photon field $(\gamma^\prime)$.  In both panels, the white (LULI) and black (PHELIX) dashed lines correspond to the expression 
$\xi_\epsilon=1$.  The left panel includes, in addition, the  exclusion regions  stemming  from various  experimental  collaborations searching 
for rotation and ellipticity in constant magnetic fields  such as BFRT \cite{Cameron:1993mr},  PVLAS \cite{Zavattini:2007ee,DellaValle:2013xs}  and Q $\&$ A \cite{Chen:2006cd}.  
The  shaded areas in the upper left corner in the right panel results from    experimental collaborations dealing with the Light Shining Through a Wall 
mechanism. The respective $95\%$  confidence  levels needed to recreate  these results are summarized  in Ref.~\cite{Ahlers:2007qf}. }
\end{figure}

The  projected exclusion regions are summarized in Fig.~\ref{fig:mb003}. They are shaded in purple and red for PHELIX and in blue and green  for LULI. These should be trustworthy as long as the limits  
lie below  the  white and black  dashed lines corresponding to $\xi_\epsilon=\epsilon m \xi/m_\epsilon=1$ for LULI and  PHELIX, respectively. In this figure, the left panel shows the  discovery potential associated 
with the pure MCP model, whereas the projected bounds including the hidden-photon effects are displayed  in the right panel. The results shown in the latter were obtained by setting  $\chi=\epsilon$, so that   
the hidden coupling constant coincides with  the natural value  $e_h=e$ [see below Eq.~(\ref{interactinglagra})].\footnote{From now on it must be understood that  the symbol $e$ refers to the absolute 
value of the electron charge.}  This assumption allows us to compare the respective outcomes  with the pure MCP model. Notice that the left panel incorporates some constraints established from other polarimetric 
searches \cite{Cameron:1993mr,Zavattini:2007ee,DellaValle:2013xs,Chen:2006cd}.  The upper bounds which result from these experiments do not represent sensitive probes of the parameter space associated 
with the model containing  the hidden-photon field \cite{Ahlers:2007rd}. Because of this fact, they are not  displayed  in the right panel. To compensate it and still put our results  into perspective, 
we include here the limits resulting from various  collaborations which deal with Light Shining Through a Wall setups \cite{Robilliard:2007bq,Cameron:1993mr,Chou:2007zzc,Ehret:2010mh}. Similar to the fermion MCPs 
model,  we observe that the most stringent sensitivity limits  appear in the vicinity of the first threshold  mass  $m_1\approx1.64\ \rm eV$. This outcome follows from a search  of the rotation 
angle. In such a situation, the projected bound turns out to be  $\epsilon< 2.3 \times 10^{-6}$ for PHELIX  and $\epsilon<7.5\times 10^{-6}$ for LULI. When comparing these results with the previously obtained 
for the model driven by fermionic MCPs [$\epsilon< 1.9 \times 10^{-6}$ for PHELIX  and $\epsilon< 6.5\times 10^{-6}$ for LULI], we note that the absence of spin degrees of freedom slightly relaxes the projected  
sensitivity. Another interesting aspect to be highlighted in Fig.~\ref{fig:mb003} is the curve shapes of the upper limits, which deviate from those coming out from the fermion MCPs model. [cp. Fig.~2 in I] 

Observe that, independently on whether the model includes paraphotons or not,  the absence of signals for PHELIX parameters leads to similar constraints.  This  fact  manifests the dominance of the  first contributions 
to the observables  in Eqs.~(\ref{rotation}) and (\ref{ellipticity}) for the given set of parameters.  We  infer that,  in the region of interest within the $(\epsilon,m_\epsilon)-$plane,  
the  characteristic times involved in the respective damping factors  $\chi^2\kappa_{\pm,1}^{-1}$  turn out to be much smaller than the pulse lengths $\tau\gg \chi^2\kappa_{\pm,1}^{-1}$. However, the behavior 
is different when the LULI parameters are used.  For masses  in the range $m_1<m_\epsilon<m_2$ the respective  upper bounds are characterized by an oscillatory  pattern whose occurence is a direct 
consequence of the photon-paraphoton  oscillations. This implies that, in such a regime, the  characteristic times $\chi^2\kappa_\pm^{-1}$  for LULI are much larger than the used  pulse lengths $\tau$;  the former 
being mainly  determined by contributions coming from  the  second threshold point $\kappa_{\pm}\simeq\kappa_{\pm,2}$ [see Eq.~(\ref{absorptioncoefficient2order})].

\begin{figure}
\includegraphics[width=6.3in]{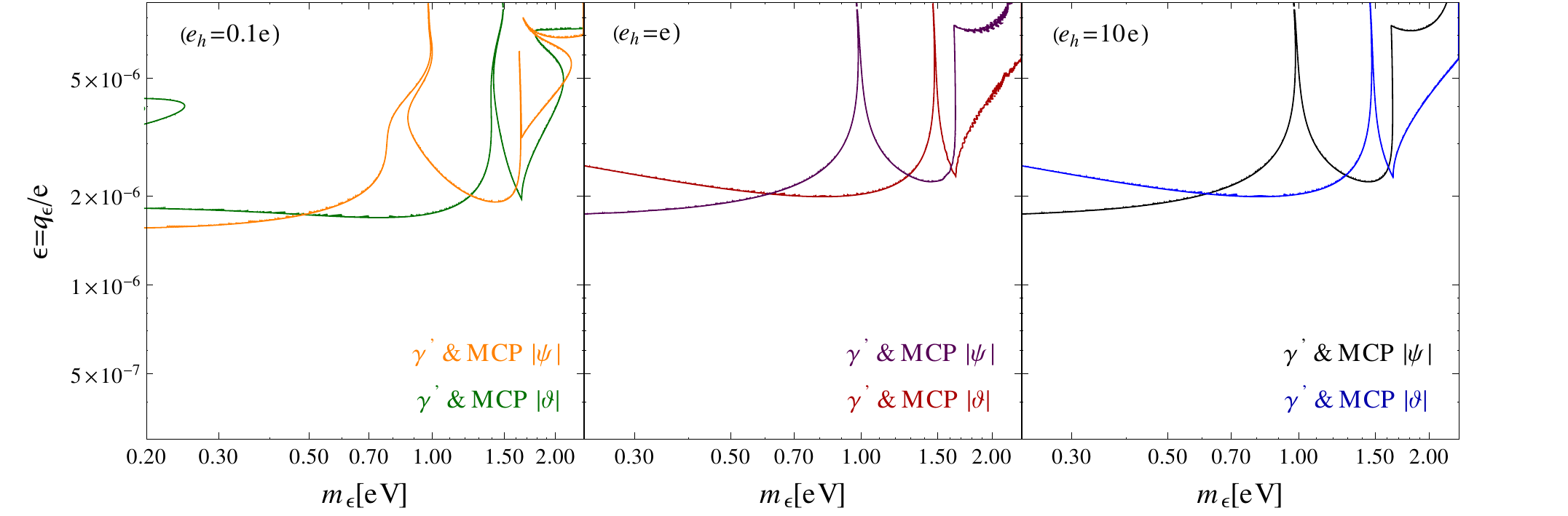}
\caption{\label{fig:mb004}Parameter space to be  ruled out for MCPs in a model with paraphotons $(\gamma^\prime)$. The  expected exclusion limits have been 
obtained by assuming the  absence of signals in a  polarimetric setup assisted by a circularly polarized wave associated with the nanosecond frontend of the  PHELIX 
laser. Here,  the projected sensitivities on the kinetic mixing parameter for various values  of the hidden coupling constant are displayed by contour lines [see legend]. }
\end{figure}

We continue our investigation by studying  the dependence of the sensitivity limits on the hidden gauge coupling  $e_h$.  Fig.~\ref{fig:mb004}  displays how the constraints for PHELIX might vary as $e_h$ changes  
by an order of magnitude around $e$. Taking the central panel [$e_h=e$] as a reference, we note that the differences between this one and the one evaluated at $e_h=10e$ [right panel] are almost imperceptible. In contrast, 
a notable distortion can be observed at $e_h=0.1e$ [left panel]. Generally speaking, both trends resemble the  results found for a spinor MCPs model. However, when directly comparing the present outcomes with those corresponding 
to the latter model [see Fig.~3 in part I of this series], we see that, at $e_h=0.1e$, the absence of spin degrees of freedom strongly modifies the qualitative behavior of the projected limits. This is not the case at $e_h=10e$,  
where the difference between the scalar and fermion models is mainly quantitative. 

Perhaps the most important conclusion that one can draw  from our results  is  that, the  sensitivity limits expected for  experiments driven by long laser pulses  of moderate intensities would allow  to  discard 
a region of the parameter  space which  has not been  excluded so far  by other  laboratory-based collaborations. Astrophysical and cosmological constraint are  stronger \cite{Jaeckel:2010ni,Ringwald:2012hr,Hewett:2012ns,Essig:2013lka}, 
though, but they  must be  considered with some care.  As we already mentioned in the introduction, the limits resulting from these scenarios strongly depend on models associated with certain phenomena which are 
not observed, such as start cooling  in the first place. The  vulnerability of these models has been addressed in various investigations and justifies the laboratory-based searches for these weakly interacting 
sub-eV particles \cite{evading,Masso:2005ym,Masso:2006gc,Jaeckel:2006id}. Uncertainties introduced by parameters such as temperature, density and microscopic energy-momentum transfer are so notable that a 
reconciliation between the astro-cosmological constraints and those resulting from the laboratory-based experiments is achievable. To put this statement in to context, let us  recall that for MCPs, a study of the 
helium-burning phase of Horizontal-Branch (HB) stars  establishes $\epsilon\leq 2\times 10^{-14}$ for $m_\epsilon\lesssim\mathrm{keV}$.  However, the lack of control on the physics occurring in such stellar objects  
might lead the omission of  suppression channels in the production of MCPs and paraphotons whose incorporation would attenuate the previous limitation.  This issue  has been analyzed carefully within the 
RM-model~\cite{Masso:2006gc}, a simple scenario in which two paraphotons--one  massless and one massive (mass $m_{\gamma^\prime}$)--are minimally coupled to dark fermions with opposite hidden charges. The 
incorporation of both paraphotons  can be  done in such a way that no additional charge labeling the elementary particles is needed and leads to $\epsilon<4 \times 10^{-8} ([\mathrm{eV}]/m_{\gamma^\prime})^2$. 
Accordingly, less severe bounds appear when the paraphoton mass $m_{\gamma^\prime}$ is getting  smaller. This fact fits very well with our approach since it relies on the fulfillment of the condition  
$m_{\gamma^{\prime}}\ll( \pi_{\pm}/\chi)^{\nicefrac{1}{2}}$ [see discussion above  Eq.~(\ref{rotation})]. Note that, at the first threshold $m_\epsilon=m_1$ resulting from PHELIX parameters,  $\chi<2.3\times10^{-6}$. 
So, the loop dominance in the photon-paraphoton oscillations is well justified as long as  $m_{\gamma^{\prime}}\ll \mathpzc{o}[0.1-1]\rm\mu eV$, for which the constraints coming from the HB stars  become much less 
stringent than the projected sensitivity estimated here. In part I of this series we explained that there are even certain sectors in $m_{\gamma^\prime}$ in which our projected upper bounds for $\chi$  turn out to 
be currently the  best model-independent  results. Similar conclusions can be drawn from  a study of a hypothetical solar emission of hidden massive photons for which the constraint  $\chi<4\times 10^{-12} (\mathrm{eV}/m_{\gamma^\prime})$ for  
$m_{\gamma^\prime}\lesssim 3\ \rm eV$ has been established \cite{Jredondo}.

%%%%%%%%%%%%%%%%%%%%%%%%%%%%%%%%%%%%%%%%%%%%%%%%%%%%%%%%%%%%%%%%%%%%%%%%%%%%%%%%%%%%%%%%%%%%%%%%%%%%%%%%%%%%%%%%%%%%%%%%%%%%%%%%%
\subsection{Characteristic of the signals in the scalar MCPs model  \label{DSM}}
%%%%%%%%%%%%%%%%%%%%%%%%%%%%%%%%%%%%%%%%%%%%%%%%%%%%%%%%%%%%%%%%%%%%%%%%%%%%%%%%%%%%%%%%%%%%%%%%%%%%%%%%%%%%%%%%%%%%%%%%%%%%%%%%%

Suppose that the outgoing probe beam acquires an ellipticity and rotation which do not coincide with the QED prediction [cp. discussion in Sec.~3.1 of I]. If their origin can be attributed to MCPs\footnote{In first instance, 
these signals migh not only result from the existence of MCPs and paraphotons but also from other light dark matter candidates such as  axion-like particles.}, the next questions of interest are: do the signals come from the 
existence of scalar or spinor MCPs; and  do they manifest the effects intrinsically associated with hidden-photons? The answers to  these questions can be obtained  by investigating  the dependencies of the observables on the 
laser parameters. In this subsection, we  provide arguments which might help to discern  the phenomenological differences that result from the various MCP models of interest. Our discussion will be based on the outcomes 
derived from  the  benchmark parameters of the nanosecond frontend of PHELIX  [standard values $\xi=6.4\times 10^{-2}$, $\tau=20\ \rm ns$, $\lambda_0=1053\ \rm nm$] and by considering a probe beam  with $\lambda=\lambda_0/2$  
which  collides head-on  with the intense laser wave. To facilitate the comparisons between the scalar and fermion models,  each figure in this subsection will encompass the same scales as used in part I of this series.  

\begin{figure}
\includegraphics[width=6 in]{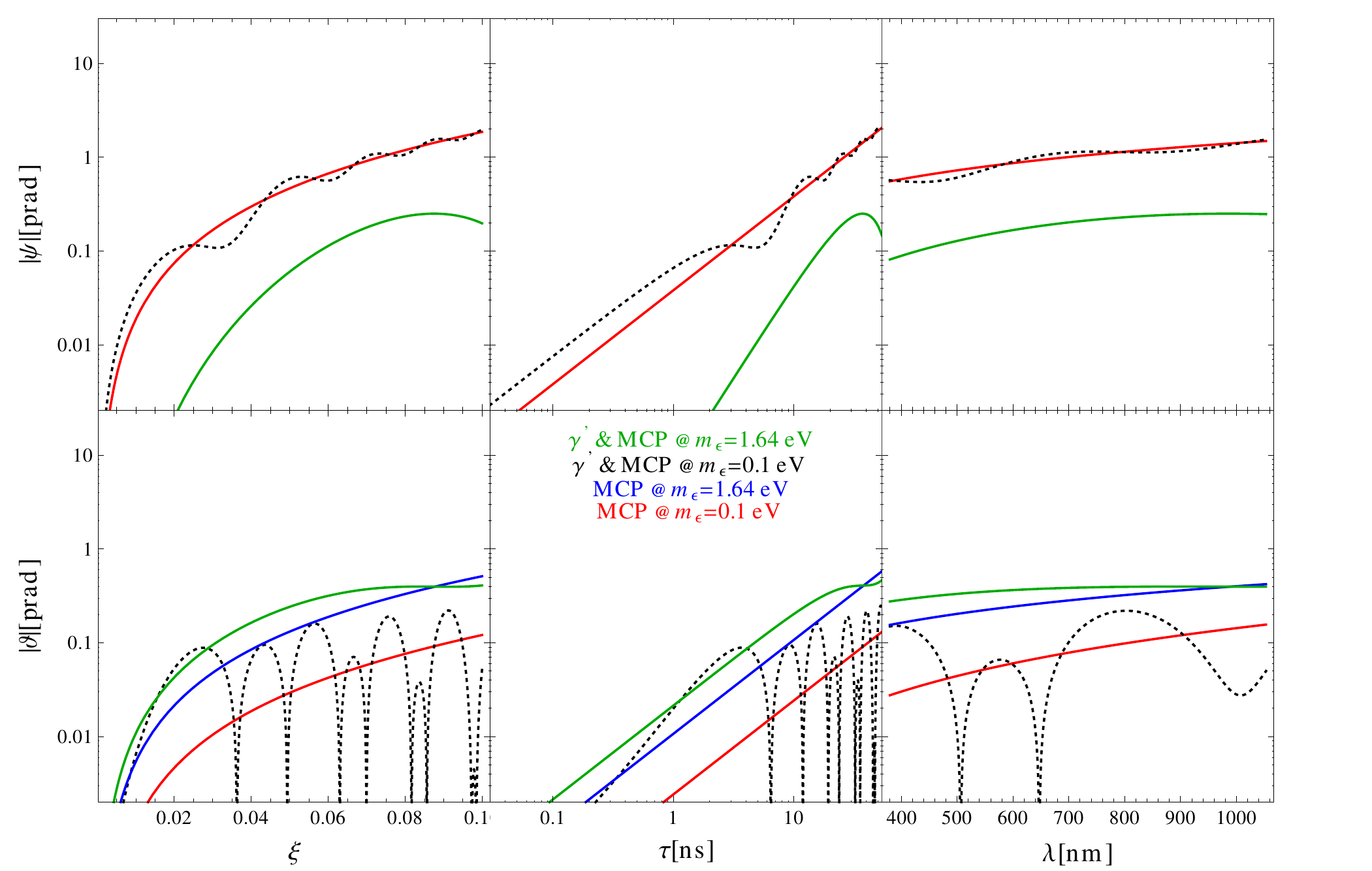}
\caption{\label{fig:mb005} Dependence of the absolute value of the ellipticity $\vert\psi\vert$ [upper panels] and rotation angle $\vert\vartheta\vert$  
[lower panels] on the  intensity parameter $\xi$ [left panel], pulse length $\tau$ [central panel] and wavelength of the probe $\lambda=\lambda_0/2$ [right panel]. As a benchmark point we assume a massless hidden 
photon field with kinetic mixing parameter $\chi=5\times 10^{-7}$ and hidden coupling $e_h=e$.  In each plot the remaining external  parameters are kept at $\xi=6.4\times 10^{-2}$, $\tau=20\ \rm ns$, $\theta=\pi$, 
and   $\lambda_0=1053\ \rm nm$  the wavelength of the intense laser field.  Here  the outcomes  resulting  from a  pure MCP model at  $m_\epsilon=0.1\ \rm eV$ are shown in red, whereas  the respective patterns at 
the first threshold mass $m_1\approx(k\varkappa/2)^{\nicefrac{1}{2}}=1.64\ \rm eV$ are  in blue. The curves in green and dotted black were obtained by  including the paraphoton field. They also correspond to the c
ase in which the  mass of the minicharges are $m_\epsilon=m_1$ and  $m_\epsilon=0.1 \ \rm eV$, respectively. Observe  that the blue curves--corresponding to the pure MCPs model at $m_1=1.64\ \rm eV$--do not appear  
in the upper panel. This is because,  at the first threshold mass,  the ellipticity  becomes  extremely tiny being  determined  by the next-to-leading order term in the absorption coefficient 
[Eqs.~(\ref{absorptioncoefficient2order})-(\ref{F3scalar})]. }
\end{figure}

\begin{figure}
\includegraphics[width=6 in]{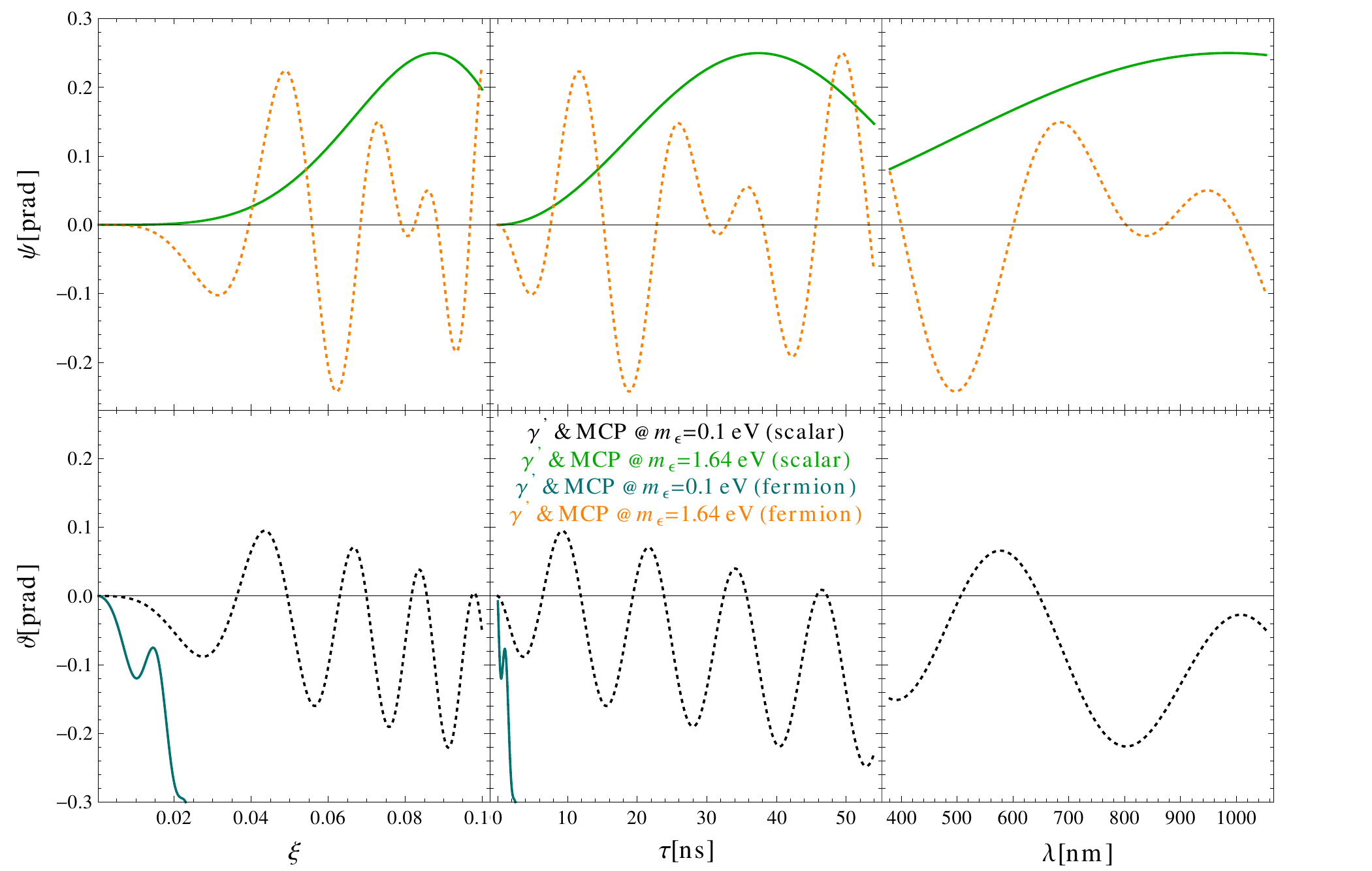}
\caption{\label{fig:mb007} Behavior of the the ellipticity $\psi(\epsilon,m_\epsilon)$ [upper panels] and rotation angle $\vartheta(\epsilon,m_\epsilon)$ [lower panels] on the  intensity parameter $\xi$ [left panel], pulse length $\tau$ [central panel] 
and wavelength of the probe $\lambda$ [right panel] in  models  including fermion versus  scalar MCPs. The same  parameters values and notation as in  Fig.~\ref{fig:mb005} are used, but the plots are in  linear scales. }
\end{figure}

The behavior of the signals with respect to the laser intensity  parameter $\xi$, the temporal length $\tau$ and the  wave length of the probe beam $\lambda=2\pi\omega^{-1}$ are displayed in Fig.~\ref{fig:mb005}. These results 
have been obtained at the first  threshold mass $m_\epsilon=m_1\approx(k\varkappa/2)^{\nicefrac{1}{2}}\approx1.64\ \rm eV$ and at  $m_\epsilon=0.1\ \rm eV$. While the outcomes associated with the pure MCP scenario are shown in blue 
and red, the results  including the paraphoton effects are shown in green and black dotted curves. To facilitate  our discussion, we will first focus on the behavior of the absolute value of the  ellipticity  $\vert\psi(\epsilon,m_\epsilon,\chi)\vert$. 
Out of the threshold, i.e. at  $m_\epsilon=0.1\ \rm eV$,  this observable [upper panels] grows with the increase of the three laser parameters. The curves associated with the model including a paraphoton field [black dotted lines in the upper panel]  manifest pronounced oscillatory  behaviors around  the 
paths followed by the pure MCPs framework [in red]. This fluctuating patterns were also revealed  in the fermion case, although in a tiny--almost imperceptible--degree only. The occurrence of this trend at $m_\epsilon=0.1\ \rm eV$  
is closely related to the photon-paraphoton oscillation, which  seems to benefit from the absence of the spin degrees of freedom.  The behavior of $\vert\psi(\epsilon,m_\epsilon,\chi)\vert$  at the first threshold mass 
[$m_\epsilon=m_1$] is different.  Here the  curves in green [upper panel]  can be exploited  to elucidate the nature of the charged carriers. To do this, we  note that the oscillatory patterns  in the ellipticity  spread 
considerably as compared with those corresponding to the fermion MCPs model [see upper panel in Fig.~\ref{fig:mb007}].  As  such,  the displayed curves for scalar MCPs  do not show  oscillations  within the investigated intervals 
for $\xi,\ \tau$ and  $\lambda$.  This fact constitutes   a remarkable property  because it implies that a slight variation of the intensity could not lead to change the signal sign for the scalar MCPs model, but  it might  
change $\psi(\epsilon,m_\epsilon,\chi)$ substantially  if  it is induced by  the fermion model. Clearly, this analysis is also applicable to the remaining parameters of the external laser wave. 

The reason why the  ellipticity curves for scalar MCPs   do not change the sign can be understood as follows:  at $m_\epsilon=m_1$, the charge carriers tend to be  produced at rest [$\mathpzc{v}_1\to 0$], so that the 
leading order terms in  the absorption coefficients [Eq.~(\ref{absorptioncoeffdecompo})] tend to vanish. As a consequence, the characteristic times $\sim\chi^2\kappa_{\pm,2}^{-1}$ increase and can reach values  much larger than the corresponding pulse length 
$\tau$. Accordingly, the exponential damping factors in Eq.~(\ref{ellipticity}) can approach unity. Besides, by quoting the refractive indices from  Ref.~\cite{Villalba-Chavez:2013gma}: $(n_i-1)\vert_{n_*=0} \approx-\alpha_\epsilon m_1^2 \xi_{\epsilon}^2/(5\pi \omega_{\pmb{k}}^2)\delta_{-,i}$ 
with $i=+,-$ we find that the  asymptotic expression for the ellipticity is determined by the oscillation probabilities between a photon and a paraphoton with negative helicities $\mathpzc{P}_{\gamma_-\to\gamma_-^\prime}$:\footnote{A general expression 
for the oscillation probability between  photon and paraphoton has already  been   derived [see Eq.~($2.38$) in I]:
$\mathpzc{P}_{\gamma_\pm\to\gamma_\pm^\prime}(\tau)\simeq \chi^2\left\{1+\exp\left(-\frac{2}{\chi^2}\kappa_\pm \tau\right)-2\exp\left(-\frac{1}{\chi^2}\kappa_\pm \tau\right)\cos\left(\frac{n_\pm-1}{\chi^2}\omega_{\pmb{k}}\tau\right)\right\}$.}
\begin{eqnarray}
\psi(\epsilon,m_1,\chi)\approx\frac{1}{4}\mathpzc{P}_{\gamma_-\to\gamma_-^\prime}\qquad\mathrm{with}\qquad \mathpzc{P}_{\gamma_-\to\gamma_-^\prime}=4\chi^2\sin^2\left(\frac{n_--1}{2\chi^2}\omega_{\pmb{k}}\tau\right).\label{approachellipatthreshold}
\end{eqnarray} Manifestly,  in Fig.~\ref{fig:mb007}, the green curves  resemble  the $\sin^2$-shape obtained above. We remark that, in contrast to the fermion model, the remaining oscillation probability in the  
scalar scenario  tends to vanish identically [$\mathpzc{P}_{\gamma_+\to\gamma_+^\prime}\approx0$].\footnote{As $\tau\ll\chi^2\kappa_+^{-1}$ and $n_+-1\approx 0$, it results  $\mathpzc{P}_{\gamma_+\to\gamma_+^\prime}(\tau)\approx0$.}  

A similar study  allows us to find  the asymptote for the absolute value of the rotation  angle $\vartheta(\epsilon,m_\epsilon,\chi)$ 
at the first threshold point [$m_\epsilon=m_1$]. In this case, 
\begin{equation}\label{approachrotationthreshold}
\vartheta(\epsilon,m_1,\chi) \approx-\frac{1}{2}\chi^2\left[\mathpzc{s}+\sin(\mathpzc{s})\right], \qquad \mathpzc{s}\equiv\frac{n_--1}{\chi^2}\omega_{\pmb{k}}\tau. 
\end{equation}Observe that, since the refractive index  $n_--1<0$, we have $\mathpzc{s}<0$ and the involved function  $\mathpzc{s}+\sin(\mathpzc{s})\leqslant0$. As a consequence,  the rotation angle does not change the sign either, 
a fact which is manifest in Fig.~\ref{fig:mb005} [lower panel]. We note that,  at  the first threshold mass [$m_1\simeq1.64\ \rm eV$], no manifestation of oscillations appears  within the range of interest in the  
external field parameters. However,  at $m_\epsilon=m_1$,  the patterns found in the fermionic  model with a  hidden photon field fluctuate about the curves which result from  the  pure MCPs scenario. At this point we shall recall that--in contrast 
to the ellipticity--such  oscillations for $\vartheta(\epsilon,m_\epsilon,\chi)$ do not change the sign [see I for details].  Therefore, if on   variating $\xi,\ \tau$ and  $\lambda$, the signal does not oscillate as described 
previously, then one could associated the measurements with the scalar model.  Still, this way of elucidating the nature of the involved charge carriers may be considered  more difficult  than the approach associated with the ellipticity 
since no change of sign arises. 

Regarding the behavior of the  rotation angle at $m_\epsilon=0.1\ \rm eV$, the occurrence of highly oscillating patterns in the model with paraphotons is notable  [black dotted curves in Figs.~\ref{fig:mb005}~and~\ref{fig:mb007}, lower panels]. The corresponding trend 
associated with  the fermion model turns out to be  much less  pronounced. While in this  last scenario there is no change of sign, in the scalar case  the signal might  change. This is because, for the present benchmark 
parameters, the characteristic time associated with the negative helicity  mode $\sim\chi^2\kappa_-^{-1}$ becomes much smaller than the pulse length  [$\tau=20\ \rm ns$], leading to an exponential suppression of the last term 
in Eq.~(\ref{rotation}). Conversely,  the characteristic time related to the  positive helicity mode is $\chi^2\kappa_+^{-1}\gg\tau$. In  such a situation, the remaining damping factor in Eq.~(\ref{rotation}) can be approached  
by unity and 
\begin{equation}\label{approachrotation}
\vert\vartheta(\epsilon,m_\epsilon,\chi)\vert \approx\frac{1}{2}\left\vert(n_+-n_-)\omega_{\pmb{k}}\tau+\chi^2\sin\left(\frac{n_+-1}{\chi^2}\omega_{\pmb{k}}\tau\right)\right\vert.
\end{equation}Thus, as in the case of the ellipticity, one might--by changing the external field parameters--use a change of sign  to elucidate whether scalar MCPs are realized or not in nature. Although 
Eq.~(\ref{approachrotation}) looks similar to Eq.~(\ref{approachrotationthreshold}) it differs from the latter in the important respect that it involves the refractive index $n_+-1$ which--in the current 
regime of mass--does not vanish identically. 

\begin{figure}
\includegraphics[width=6 in]{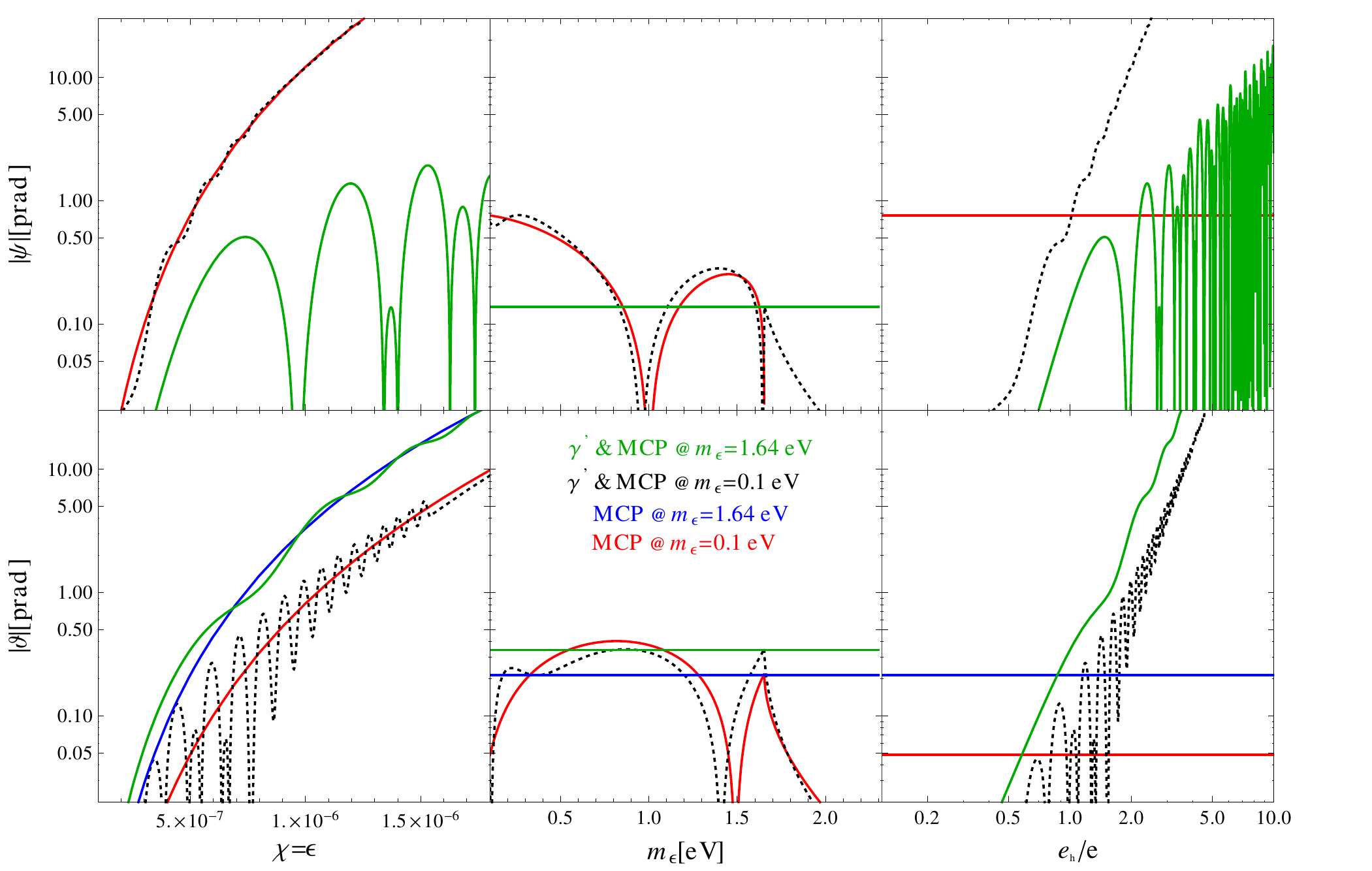}
\caption{\label{fig:mb006} Dependence of the absolute value of the ellipticity $\vert\psi\vert$ [upper panels] and rotation angle $\vert\vartheta\vert$  
[lower panels] on the  kinetic mixing parameter $\chi$ [left panel], mass $m_\epsilon$ [central panel] and the relative hidden coupling $e_h/e$ [right panel]. 
The same  benchmark values and notation as in Fig.~\ref{fig:mb005} are used.}
\end{figure}

Finally, in Fig.~\ref{fig:mb006}, the dependencies of the ellipticity and rotation of the polarization plane with respect to some  unknown parameters are shown. The vertical central panel of this figure  displays how  the signals 
might change  with the mass $m_\epsilon$ of this hypothetical charge carriers. As in the fermion model, the ellipticity resulting from  the  scenario without paraphotons  reveals a discontinuity at the first threshold mass  
[red curve], discussed in Sec.~\ref{spectraldecompos}, which is smoothed as soon as a hidden photon field is taken into account [dotted black curve].  As a sight remark, we point out that  at the first threshold, the 
ellipticity  is  constant in  both models.  Note that the blue curves--corresponding to the pure MCPs model at $m_1=1.64\ \rm eV$--do not appear  in the upper panels neither in Fig.~\ref{fig:mb005}  nor in Fig.~\ref{fig:mb006}.
This is because,  at the first threshold mass,  the ellipticity  becomes  extremely tiny being  determined  by the next-to-leading order term in the absorption coefficient [Eqs.~(\ref{absorptioncoefficient2order})-(\ref{F3scalar})]. 
We note that, in contrast to the ellipticity,  the dependence of $\vert\vartheta(\epsilon,m_\epsilon,\chi)\vert$ with respect to the mass  $m_\epsilon$ follows a continuous paths in both models. 
Regarding  the left and right  vertical panels, they  illustrate   how  both observables depend on the  mixing parameter $\chi$ and the relative hidden coupling $e_h/e$. In both panels  the fluctuating  patterns 
for the ellipticity [Eqs.~(\ref{approachellipatthreshold})] and rotation of the polarization plane [Eq.~(\ref{approachrotation})],  at the respective masses $m_1=1.64\ \rm eV$  and  $m_\epsilon=0.1\ \rm eV$  can be seen.   
Particularly, the outcomes associated with the latter observable in the lower left panel manifest that the curve including a hidden-photon field is modulated around the  pure MCPs contribution  [first term in the right-hand 
side of [Eq.~(\ref{approachrotation})]. Both panels show   a  fast decrease of  the observables  for small values of $\chi$, a  trend  which is  also  manifest  with respect to $e_h/e$ [black dotted curve]. We remark that, 
in the right panel, the outcomes resulting from  the pure MCP scenario [horizontal red and blue lines] are not sensitive to  variations  of the relative hidden coupling because the latter only emerges within  the framework of a 
hidden-photon field.

%%%%%%%%%%%%%%%%%%%%%%%%%%%%%%%%%%%%%%%%%%%%%%%%%%%%%%%%%%%%%%%%%%%%%%%%%%%%%%%%
\section{Conclusions and outlook}
%%%%%%%%%%%%%%%%%%%%%%%%%%%%%%%%%%%%%%%%%%%%%%%%%%%%%%%%%%%%%%%%%%%%%%%%%%%%%%%%

Experiments designed to detect the QED  vacuum birefringence in laser pulses might provide  insights about   light dark matter candidates such as MCPs and paraphotons. Throughout 
this investigation, we have paid special attention to the capability which long laser pulses [$\tau\sim \rm ns$] of moderate intensities [$\xi<1$] offer for the exploration of new domains 
of particle physics. Particularly, we have pointed out that their long durations  compensate the small intensities associated with them and  the combination of this feature with the fact 
that they  are also characterized by a well-defined frequency manifests the realization of thresholds in which the projected sensitivities can be higher than those achieved in experiments driven 
by dipole magnets. We have noted that--depending on the external parameters--the absence of spin can facilitate or counteract the photon-paraphoton oscillations, as compared with the fermion 
MCPs model. This intrinsic property might manifest through the probe photon beam and, can be exploited to discern the quantum statistics of these particle candidates. 
A special emphasis has been laid on a plausible change in the ellipticity sign that the probe photon can undergo, depending  upon the MCPs nature. 

Finally, we emphasize that the  treatment used in this investigation is valid only for $\xi_\epsilon\ll1$. It would be interesting to extend the present research to the case in which 
$\xi_\epsilon>1$.  We remark that, the estimated upper bounds  [$\epsilon\sim 10^{-6}-10^{-5}$ for  $m_\epsilon\sim 0.1-1\ \rm eV$] can   lead to an intensity  parameter greater 
than unity  [$\xi=\frac{m_\epsilon}{\epsilon m}\xi_\epsilon\gg1$],  provided $\xi_\epsilon\gg 1$. Corresponding laser sources exist. Indeed, intensities as large as $\sim10^{22}\ \rm W/cm^2$ 
have already been  achieved by the  HERCULES petawatt system \cite{yanovsky} and a substantial intensity  upgrade is foreseen at  ELI  and  XCELS \cite{ELI,xcels}. In connection with these high-intensity 
petawatt source,  the HIBEF consortium \cite{HIBEF}  has proposed an experiment  to measure vacuum birefringence for the first time by combining a very intense optical pulse with  $\xi\gg1$  and a probe x-ray 
free electron laser \cite{Heinzl}. Certainly,  these measurements will provide a genuine opportunity to  search for axion-like particles, MCPs and paraphotons. However, in constrast to 
our treatment, a  theoretical description of a polarimetric experiment assisted by such  pulses  is  complicated by the fact that--as a result of the focusing--their typical spatial extensions  
$\mathpzc{d}\sim \rm \mu m$  are comparable with their  wavelengths.  As a consequence, the monochromatic model for the external field [Eq.~(\ref{externalF})] is no longer valid  
and the  pulse profile becomes relevant for the  establishment of the exclusion limits. For axion-like particles a study of this nature has  already been carried out \cite{Villalba-Chavez:2013bda}, 
but it remains intriguing to see  how the wave profile can influence the upper bounds associated with MCPs and hidden photon fields. 

%%%%%%%%%%%%%%%%%%%%%%%%%%%%%%%%%%%%%%%%%%%%%%%%%%%%%%%%%%%%%%%%%%%%%%%%%%%%%%%%
\section*{Acknowledgement}
%%%%%%%%%%%%%%%%%%%%%%%%%%%%%%%%%%%%%%%%%%%%%%%%%%%%%%%%%%%%%%%%%%%%%%%%%%%%%%%%

We gratefully acknowledge useful discussions with  Holger Gies and funding by the German Research Foundation (DFG) under Grant No. MU 3149/2-1.

\end{document}